\documentclass[amsmath,amssymb,reqno,tbtags,psamsfonts,10pt,a4paper,twocolumn,superscriptaddress,nobalancelastpage,prx]{revtex4-2}
\makeatletter
\renewcommand{\p@subsection}{}
\renewcommand{\p@subsubsection}{}
\makeatother

\usepackage[english]{babel}
\usepackage{stmaryrd}
\usepackage{color,graphicx}
\usepackage{upgreek}
\usepackage{dcolumn}% Align table columns on decimal point
\usepackage{bm}% bold math
\usepackage[per-mode=symbol]{siunitx}

\usepackage{array}

\usepackage[colorlinks=true, linkcolor=blue]{hyperref} % turns the page numbers in the index into links, among other things

\usepackage{cleveref}

\newcommand\eps{\ensuremath{\varepsilon}}

\def\kp{k}
\def\kd{\kappa}
\def\Lp{K}
\def\betaGhost{}

\usepackage{color}

\def\gsim{\mathrel{\raise.3ex\hbox{$>$\kern-.75em\lower1ex\hbox{$\sim$}}}}
\def\lsim{\mathrel{\raise.3ex\hbox{$<$\kern-.75em\lower1ex\hbox{$\sim$}}}}

\makeatletter
\let\saved@includegraphics\includegraphics
\AtBeginDocument{\let\includegraphics\saved@includegraphics}
\renewenvironment*{figure}{\@float{figure}}{\end@float}
\makeatother

\begin{document}

\title{Nash epidemics}
\bigskip
\author{Simon K. Schnyder}
\email[]{skschnyder@gmail.com}
\affiliation{Institute of Industrial Science, The University of Tokyo,
4-6-1 Komaba, Meguro-ku, Tokyo 153-8505, Japan}
\author{John J. Molina}
\affiliation{Department of Chemical Engineering, Kyoto University, Kyoto 615-8510, Japan}
\author{Ryoichi Yamamoto}
\affiliation{Department of Chemical Engineering, Kyoto University, Kyoto 615-8510, Japan}
\author{Matthew S. Turner}
\email[]{m.s.turner@warwick.ac.uk}
\affiliation{Department of Physics, University of Warwick, Coventry CV4 7AL, UK}
\affiliation{Institute for Global Pandemic Planning, University of Warwick, Coventry CV4 7AL, UK}
% \affiliation{Department of Chemical Engineering, Kyoto University, Kyoto 615-8510, Japan}

\date{\today}% It is always \today, today,

\begin{abstract}
Faced with a dangerous epidemic humans will spontaneously social distance to reduce their risk of infection at a socio-economic cost.
Compartmentalised epidemic models have been extended to include this endogenous decision making: 
Individuals choose their behaviour to optimise a utility function, self-consistently giving rise to population behaviour.
Here we study the properties of the resulting Nash equilibria, in which no member of the population can gain an advantage by unilaterally adopting different behaviour.
We leverage a new analytic solution to obtain,
(1) a simple relationship between rational social distancing behaviour and the current number of infections; 
(2) new scaling results for how the infection peak and number of total cases depend on the cost of contracting the disease; 
(3) characteristic infection costs that divide regimes of strong and weak behavioural response and depend only on the basic reproduction number of the disease; 
(4) a closed form expression for the value of the utility.
We discuss how these analytic results provide a deep and intuitive understanding into the disease dynamics, 
useful for both individuals and policymakers. 
In particular the relationship between social distancing and infections represents a heuristic that could be communicated to the population to encourage, or ``bootstrap'', rational behaviour.
\end{abstract} % abstract must not exceed 200 words.

\maketitle

% \section*{Significance Statement}
% We find several previously unknown analytic relationships between the degree of rational social distancing and disease dynamics. The linear relationship between the degree of rational social distancing and the current number of infections represents a simple, broadly accessible, heuristic that could be communicated to the population in order to encourage, or ``bootstrap'', rational behaviour. We are also able to show that a single characteristic infection cost emerges that depends only on $R_0$ and find new scaling results for the peak of the infections and herd immunity level.

% \section{Introduction}

Throughout history, epidemics caused by infectious diseases have caused considerable harm to humans. Individuals
are typically assumed to be able to adjust their behaviour in reaction to the threat of an epidemic \cite{Reluga2010,Fenichel2011,Wang2016,Chang2020,Verelst2016,Bhattacharyya2019,Makris2020,Yan2021,Reluga2011,Lux2021}.
In order to make choices about their behaviour individuals can weigh the costs and benefits of the outcomes of their behaviour. They may reduce their social activity when infections are high, in order to reduce the probability of becoming infected themselves, provided that the health costs outweigh the social and economic costs. A common assumption is that individual agents act rationally, i.e. to maximise an objective function or economic utility. 
This remains one of the fundamental assumptions of modern economic theory despite its limitations~\cite{kahneman2003maps}.
Rational individuals, who aim to maximise their individual objective function, end up targeting a Nash equilibrium~\cite{Reluga2010,Reluga2011,Fenichel2011,McAdams2020,Eichenbaum2021} rather than the global utility maximum, which requires a coordinated effort to maximise a collective objective function
~\cite{Rowthorn2020,Toxvaerd2020,Makris2020,Li2017}. 
Our work directly builds on this approach. It is possible to bring a Nash equilibrium into alignment with the global optimum~\cite{Rowthorn2020,Bethune2020,Aurell2022,Schnyder2023b}, e.g. via tax and subsidy incentives~\cite{Althouse2010} which can be designed to bias rational individual behaviour appropriately.

While these studies tend to employ highly stylised mean-field compartmentalised models, they demonstrate the feasibility of such approaches.
Such models can be extended to more accurately represent the complexity of epidemics and the systems in which they occur, such as
additional compartment types with different risk and behaviour profiles \cite{Acemoglu2020,Fenichel2011,Prem2017,Huang2022,Tildesley2022,Keeling2022}, seasonal effects \cite{He2013}, waning immunity \cite{Giannitsarou2021,Keeling2022} e.g. due to new variants \cite{Schwarzendahl2022}, as well as spatial, transmission or behavioural heterogeneity~\cite{Mossong2008,Tildesley2010,Prem2017,Sun2021,Hill2023}. Other approaches feature spatial \cite{Chandrasekhar2021} and temporal networks \cite{Holme2012,Holme2015}, and/or agent-based models \cite{Ferguson2006,Tanimoto2018,Mellacher2020,Grauer2020}. 
Others have worked to incorporate uncertainty and noise, by considering stochastic control \cite{yong1999stochastic,Lorch2018,Tottori2022,Tottori2023b,Tottori2023}, 
decision making under uncertainty \cite{Barnett2023,Shea2023} and by understanding the robustness of control \cite{Kantner2020,Kohler2021,Morris2021}. There have also been studies on inferring model structure and epidemiological properties from observed data~\cite{Adhikari2020,Pietzonka2021,Li2021,Molina2022}.  Finally, we also remark on the intriguing possibility of allowing individual opinions to directly influence policy makers \cite{Mellacher2023}.

Nash equilibria are widely believed to occur within such idealised models that incorporate endogenous behaviour during epidemics. However, until this work, solutions have only been accessible numerically. This is because the problem is intrinsically nonlinear, both at the level of the epidemiological dynamics and the objective function, leading to nonlinear control equations.  Here, we provide for the first time, an analytic solution to the nonlinear time-dependent equilibrium control equations. This also demonstrates the existence of such a Nash equilibrium.  In the limit of vanishing infection cost our results trivially recover the known analytic solutions for compartment models with constant basic reproduction number \cite{Kermack1927,Miller2012,Harko2014,Miller2017,Kroger2020}, i.e. without endogenous rational behaviour.

We focus on the case where the cost of infection is constant and where the government takes no role in directing the response to the epidemic. This situation has been already discussed, e.g. by \cite{Reluga2010} among many others, but only using numerical solutions. We do not investigate other possible policy interventions such as vaccination and treatment strategies, \cite{Bauch2003,Bauch2004,Reluga2006,Tildesley2006,Reluga2011,Chen2014,Wang2016,Tanimoto2018,Chang2020,Toxvaerd2020,Grauer2020,Moore2021b,Moore2022,Hill2022,Keeling2022,Keeling2023}, or isolation, testing, and active case-tracing strategies \cite{Kucharski2020,Piguillem2020}. We also ignore the situation where a vaccine becomes available during the epidemic. While the early arrival of a vaccine would have consequences for both equilibrium and globally optimal behaviour \cite{Reluga2010,Makris2020, Eichenbaum2021,Schnyder2023}, this lies outside of the scope of this work.

\section{Epidemic dynamics}

We use a standard SIR compartmentalised model \cite{Kermack1927} for the epidemic. The population is divided into {\underline s}usceptible, {\underline i}nfected and {\underline r}ecovered compartments, the latter implicitly including fatalities. The compartments evolve over time as
\begin{align}
\dot s&=-\kp\>s\>i\cr
\dot i&=\kp\>s\>i-i \label{dimlessgeneral} \\
\dot r&=i
\nonumber
\end{align}
Here a dot denotes a time derivative and the time dependence of $s(t)$, $i(t)$, $r(t)$ and $k(t)$ is omitted for brevity. We normalise the compartments, $1 = s + i + r$.
We use one timescale for both recovery and the duration of infectiousness, for simplicity, and have rescaled the equations so that time $t$ is measured in units of this single timescale.  
The initial conditions are set as $s(0) = s_0$, $i(0) = i_0$, $r(0) = r_0$, with $s_0, i_0, r_0 \geq 0$ and $s_0 + i_0 + r_0 = 1$.
In all figures we arbitrarily select a time origin $t=0$ where the epidemic is in its very early stages according to 
$r(0)=i(0)/(R_0-1)=10^{-6}$ with $s(0) = 1 - i(0)-r(0)$.

The population's average social activity behaviour is encoded in the current infection rate, assumed to satisfy $k(t) \geq 0$ although our analytic results later suggest a stronger bound $k(t) \geq 1$. 
We assume that the disease exhibits a natural level of activity in the absence of any behavioural modification that is a constant known as the basic reproduction number $R_0$. Below we use the case $k(t)=R_0$ to establish a non-behavioural baseline dynamics for comparison. 

\section{Nash equilibrium behaviour}
\label{sec:Hamiltonian_Nash}

In order to study self-organised behaviour, we imagine an average individual making decisions about their own behaviour. This represents a mean-field game \cite{Bensoussan,Carmona}, for the Nash equilibrium of which a set of ordinary differential equations can be straightforwardly derived \cite{OptimalControlBook,Reluga2011}.

The individual's effect on the epidemic is negligible but they can influence their own fate by selecting a strategy $\kd(t) \geq 0$ which it is initially assumed can differ from the population-averaged strategy $\kp(t)$. 
The probabilities that an individual is in each of the compartments evolves over time according to
\begin{align}
\dot \psi_s&=- \kd\psi_s i\cr
\dot \psi_i&= \kd\psi_s i-\psi_i
\label{dimlesspsigeneral}
\end{align}
Lowering $\kd(t)$ directly increases the probability of the individual remaining susceptible and reduces their probability of becoming infectious.
While these equations are similar to eqs.~\ref{dimlessgeneral}, they couple to the infectious compartment of the population $i$ as the only donor of infection. 

We assume that an individual has rational interests that can be captured by an objective function or utility. In general this will depend on both their own and the population behaviours, $U(\kappa(t), k(t))$. The individual  seek to maximise this objective function. Assuming that the population consist of identical individuals, a Nash equilibrium exists if there is a strategy $k=\kappa(t)$, adopted by the population, and the individual cannot improve their outcome by unilaterally deviating from the behaviour $\kappa$,
\begin{align}
	U(\tilde \kappa(t), \kappa(t)) \leq U(\kappa(t), \kappa(t))\quad
	 \text{ for any } \tilde\kappa(t).
\label{eq:nash_definition}	 
\end{align}
In order to find this Nash strategy one first maximises  $U(\kappa, k)$ over $\kd$ for an arbitrary, exogenous $\kp$ \cite{Reluga2011}. This constitutes a standard constrained optimisation problem. To make the strategy self-consistent, one then assumes that all individuals in the population would optimise their behaviour in the same way, and therefore $\kp=\kd$. This then automatically results in $\psi_s = s$, $\psi_i = i$ with dynamics that corresponds to the Nash equilibrium.

In this work, we focus on an idealised individual objective function or utility $U$ with, see ref. \citenum{Schnyder2023b},
\begin{align}
U =&\ \int_{0}^{t_f} u(t) dt + U_f  \label{eq:U_with_salvage}\\
u =&\ f^{-t}\left[-\alpha\> \psi_i -\beta\>(\kd -R_0)^2\right]
\label{Uda}
\end{align}
The average infection cost is given by $\alpha$ (this also includes the cost of death) with $\alpha\geq 0$.

 The social and financial costs of social distancing are parametrised by a constant $\beta>0$. In what follows we choose to work in units of utility in which $\beta = 1$, without loss of generality. The quadratic form of this social distancing term encodes that it is costly to deviate from one's default behaviour and ensures that  an individual would naturally select behaviour corresponding to $\kappa = R_0$ if there were no epidemic (or it bore no cost). 
Motivated by the observation that no individual, regardless of compartment, can socially distance without incurring a cost, we assume that the cost of social distancing always applies, in contrast to other work, e.g. \cite{Reluga2010, Toxvaerd2020} where the cost of social distancing is paid mostly by the $s$-compartment. Our choice corresponds to an approximation in which individuals are uncertain about which compartment they find themselves in, e.g. when infections can be asymptomatic. 
 
In general one might truncate the utility integral at a final time $t_f$. One approach here is to assuming that at that time the susceptible compartment becomes completely and perfectly vaccinated. The cost of being infected when the vaccine becomes available is expressed as a so-called salvage term $U_f =\ -f^{-t_f}\alpha\ \psi_{i}(t_f)/(1+\ln f)$, see \cref{sec:salvage_term} for a derivation. If $t_f$ is comparable to or shorter than the duration of the epidemic then it can have a qualitative effect on rational decision-making 
and thus the course of the epidemic \cite{Reluga2010,Makris2020,Schnyder2023}. However, this lies outside of the scope of this work, and so we choose $t_f\to\infty$. In this case, 
vaccination plays no role in decision making and $i(t_f)\to 0$.

Since the utility function is convex, we expect that the optimisation problem has a (unique) solution. We directly demonstrate uniqueness and existence by calculating the analytic solution to this problem below.

We use a standard Hamiltonian/Lagrangian approach \cite{OptimalControlBook,Reluga2011}, which in optimal control theory is referred to as Pontryagin's maximum principle \cite{Pontryagin}, 
to calculate the optimal behaviour of an individual $\kd$ in response to an exogenous behaviour $\kp$ and the corresponding course of the epidemic. 
This approach allows for reformulating the optimisation problem as a boundary value problem which is generated from an auxiliary function, the Hamiltonian.  The Hamiltonian for the individual can be expressed by
\begin{align}
	H  
	=&\ u + v_s (-\kd\psi_s i) + v_i (\kd\psi_s i-\psi_i) \cr
	= &-f^{-t}\left[\alpha \psi_i +\betaGhost\,(\kd -R_0)^2\right] \cr
	  &- (v_s - v_i) \kappa \psi_s i - v_i \psi_i
\end{align}
The Lagrange fields $v_s(t)$ and $v_i(t)$, expressing the expected present value of the utility of being in each compartment at each point in time \cite{Reluga2011} enforce the constraint of the dynamics to eqs.~(\ref{dimlesspsigeneral}). 
Their equations of motion are
\begin{align}
\dot v_s &= -\frac{\partial H}{\partial \psi_s} =  (v_s - v_i) \kappa i	
\label{eq:individual_optimal_values_s}
\\
\dot v_i &= -\frac{\partial H}{\partial \psi_i} = f^{-t} \alpha + v_i
\label{eq:individual_optimal_values_i}
\end{align}
with boundary conditions
\begin{align}
	v_s(t_f)	= \frac{\partial U_f}{\partial \psi_{s,f}} = 0 
	,\ 
v_i(t_f) = \frac{\partial U_f}{\partial \psi_{i,f}} = \frac{-f^{-t_f}\alpha}{1 + \ln f}.
\label{eq:individual_values_bc}
\end{align} 
Given the exogenous course of the epidemic in the population, the individual can optimise their own utility be choosing the strategy that satisfies $0 = \partial H/\partial \kappa$. From this we obtain
$	\kappa = R_0 - \frac{f^t}{2\betaGhost} (v_s - v_i) \psi_s i$.
Assuming that the population consist entirely of identical individuals who would all independently from each other choose the same strategy, we can conclude that the average population behaviour must be self-consistently given by $k(t) = \kappa(t)$. Hence, this gives rise to a Nash equilibrium. Then, naturally also $s = \psi_s$ and $i = \psi_i$, and
\begin{align}
	k = \kappa = R_0 - \frac{f^t}{2\betaGhost} (v_s - v_i) s i.
	\label{eq:individual_optimal_kappa}
\end{align}
The variational approach as stated only yields conditions sufficient to identify extrema. To confirm that a solution is a maximum, $\partial^2 H/\partial \kappa^2 < 0$ is required. Here we find $\partial^2 H/\partial \kappa^2  = -2 < 0$.

\section{Analytic solution}

The Nash equilibrium $k(t)$ optimising the utility 
\cref{eq:U_with_salvage}
is given by the solution of \cref{dimlessgeneral},
\cref{eq:individual_optimal_values_s,eq:individual_optimal_values_i,eq:individual_values_bc}, in conjunction with the optimality condition \cref{eq:individual_optimal_kappa}.
From here, we calculate the analytic solution for this set of equations. We assume the case without economic discounting, $f = 1$.

Firstly, we work with the integrated fraction of infected cases up to time $t$, i.e. the fraction of recovered cases $r$, defined as 
\begin{equation}
r=\int_{0}^t i(t')dt' + r_0
\label{r}
\end{equation}
noting that $i=\dot r$. In what follows it is convenient to consider an implicit form for the behaviour $\kp(r)$. Because $r(t)$ is monotonic we can rely on a one-to-one mapping between $\kp(t)$ and $\kp(r)$. 
The second transformation involves defining
$\dot \Lp=\kp\>i$, 
hence $\Lp$ obeys
\begin{equation}
\Lp(r)=\int_{r_0}^r\kp(r')dr'
\label{L}
\end{equation}
Equations (\ref{dimlessgeneral}) then lead to $\dot s=-s\dot\Lp$ 
which integrates to
\begin{equation}
s = s_0 e^{-\Lp(r)}
\label{ssoln}
\end{equation}
Using $1=s+i+r$, we obtain directly
\begin{equation}
i=1-r-s_0e^{-\Lp(r)}
\label{isoln}
\end{equation}
Since $i = \dot r$, we can integrate this equation to obtain
\begin{equation}
F(r)\equiv\int_{r_0}^r\frac{dr'}{1-r'-s_0e^{-\Lp(r')}} = t \ \ \ {\rm hence} \ \  r=F^{-1}(t)
\label{isolnp}
\end{equation}

Recalling our assumption that $t_f\to \infty$, we can conclude that $i(t_f)\to 0$.
In this limit the cumulative total of infections reaches its final value given by the non-zero root of
\begin{equation}
r_f+s_0e^{-\Lp(r_f)}=1
\label{ELBCc}
\end{equation}
\Cref{r,L,ssoln,isoln,isolnp,ELBCc} above hold irrespective of the form of the objective function. 

\begin{figure}[tbp]
  \centering \includegraphics[width=\columnwidth]{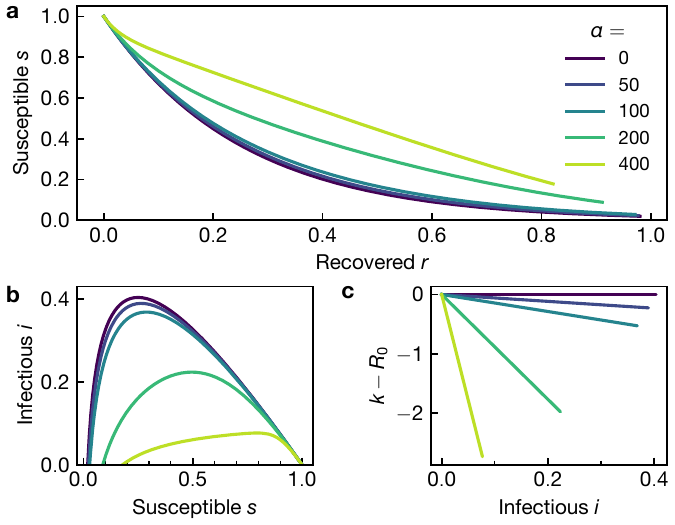}
	  \caption{
	\textbf{Direct plots of the analytic solution.} 
	(a) The analytic solution of the Nash equilibrium social distancing problem as obtained in \cref{eq:analytic_solution} as a function of the recovered $r$ for an exemplary range of infection costs $\alpha$ and $R_0 = 4$. 
Initial conditions here and in all following figures are set to $r_0 = 10^{-6}$ and $i_0 = 3\cdot 10^{-6}$.
	(b) The fraction of infectious $i$ as a function of the susceptible $s$ for the same range of $\alpha$.
	(c) Deviation of the social distancing behaviour $k$ from the pre-epidemic default $R_0$ as a function of $i$, emphasising their linear relationship as established in \cref{eq:k_vs_i}.
	}
	   \label{fig:Fig2}
\end{figure}

Concerning the Lagrange fields, we can see directly from \cref{eq:individual_optimal_values_i} that $v_i(t) = -\alpha$, whereas
 $v_s$ follows from \cref{eq:individual_optimal_values_s}
\begin{gather}
	\dot v_s = (v_s + \alpha) \dot \Lp
\end{gather}
Integrating we obtain
\begin{gather}
v_s + \alpha = \mu e^{\Lp}	= \mu \frac{s_0}{s}
\end{gather}
with a constant $\mu$. 
From the boundary condition $v_s(t_f) = 0$ we can conclude
\begin{gather}
	v_s  = \alpha \frac{s_f}{s} - \alpha
\end{gather}
with $s_f = s(t_f)$. 
The optimal behaviour is then given by \cref{eq:individual_optimal_kappa}
\begin{align}
	k &
	= R_0 - \frac{\alpha s_f}{2}  i
 	\label{eq:k_vs_i}
\end{align}
This is tremendously simple: the equilibrium strength of social distancing $k-R_0$ is  proportional to both the number of infectious cases and the cost of infection at any given time, see \cref{fig:Fig2}c.
With $s = s_0e^{-\Lp(r)}$ we have
 \begin{align}
 \partial_r s = -s \partial_r K = - s k
 \label{eq:dsdr}
 \end{align}
and therefore, inserting \cref{eq:k_vs_i} and $i = 1-r-s$,
\begin{align}
\partial_r s &= -s [R_0 - \frac{\alpha s_f}{2}(1 - r - s)] \cr
&= -s [a + br + bs]
\label{eq:beautiful}
\end{align}
with $a = R_0 - b$ and $b = \alpha s_f/2$.
This has an analytic solution that satisfies $s(r \to r_0) \to s_0$
\begin{align}
	s(r) = \frac{\exp\left[- \frac{1}{2}(r-r_0)(2 a + b(r+r_0))\right]}
	{\frac{1}{s_0} - \sqrt{\frac{\pi b}{2}} \exp\left[\frac{(a+br_0)^2}{2b}\right]\left(\text{Erf}\left[\frac{a+br_0}{\sqrt{2b}}\right] - \text{Erf}\left[\frac{a + br}{\sqrt{2b}}\right]\right)}
	\label{eq:analytic_solution}
\end{align}
Using $r_f = 1-s(r_f)$, we can self-consistently determine $r_f$ and thus obtain the solution. 
We show the result of \cref{eq:analytic_solution} for a range of infection costs $\alpha$ in \cref{fig:Fig2}a.
The analytic solution for the infectious compartment $i(r) = 1 - r - s(r)$ can be plotted in a natural way on the s-i plane, see \cref{fig:Fig2}b. 

\begin{figure}[tbp]
  \centering \includegraphics[width=\columnwidth]{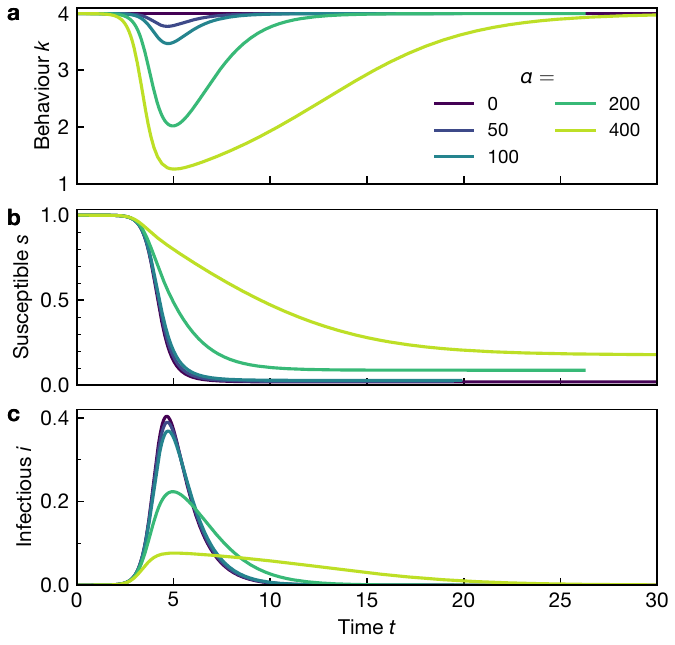}
	  \caption{
	\textbf{Analytic solution as a function of time.} (a) Equilibrium social activity behaviour of the population $k(t)$ and corresponding dynamics of the disease (b) $s$ and (c) $i$ for an exemplary range of infection costs $\alpha$ and $R_0 = 4$. 
 Since infections incur a cost, the equilibrium behaviour seeks to avoid excessive infections by self-organised social distancing. The higher the cost, the more reduced social activity $k$ becomes. 
 }
	   \label{fig:Fig1}
\end{figure}

In our approach, time is parametrised as \cref{isolnp},
which can easily be evaluated numerically. The analytic solution can then be plotted in the typical way, \cref{fig:Fig1}.

\begin{figure*}[ht!]
  \centering \includegraphics[width=1.99\columnwidth]{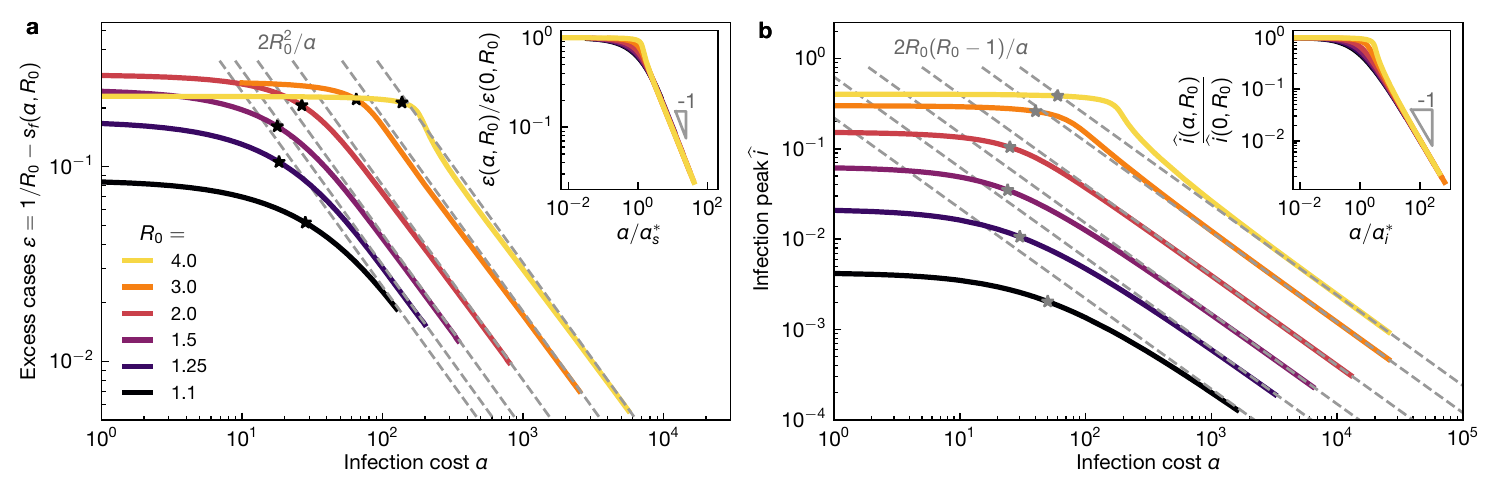}
	  \caption{
	\textbf{Scaling.}
	(a) Excess cases $\eps(\alpha,R_0)$ vs. infection cost $\alpha$ for a range of basic reproduction numbers $R_0$. The high infection cost asymptotes, see \cref{eq:excess_cases_high_alpha}, are shown as dashed lines and the crossover costs $\alpha^\star_s$, see \cref{eq:excess_cases_crossover}, as black stars.
Inset:	The data collapses onto the low-$\alpha$ and high infection cost asymptotes by rescaling the cost $\alpha$ with the crossover cost $\alpha_s^\star$, see \cref{eq:excess_cases_crossover}, while rescaling $\eps(\alpha,R_0)$ with its non-behavioural limit, see \cref{eq:excess_cases_nonbehavioural}.
	(b) The infection peak $\hat i$ vs. $\alpha$ for a range of $R_0$. The high infection cost asymptotes, see \cref{eq:peak_high_alpha}, are shown as dashed lines and the crossover costs $\alpha^\star_i$, see \cref{eq:peaks_crossover}, as grey stars.
Inset:	The data collapses onto the low-$\alpha$ and high infection cost asymptotes by rescaling the cost $\alpha$ with the crossover cost $\alpha^\star_i$, \cref{eq:peaks_crossover}, while rescaling the peak height with its non-behavioural limit, see \cref{eq:peak_nonbehavioural}.
	}
  \label{fig:excess_cases}
  \label{fig:infection_peak}
\end{figure*}

\begin{figure}[ht!]
  \centering \includegraphics[width=\columnwidth]{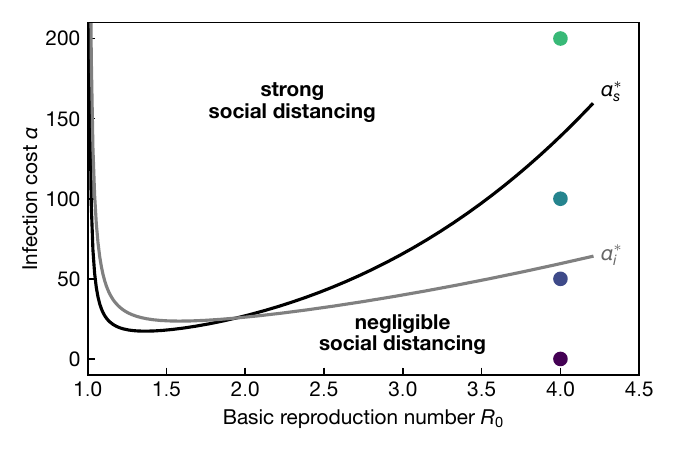}
	  \caption{
	\textbf{Behavioural response.} 
Characterisation of the Nash equilibrium response in the $R_0$ -- $\alpha$ parameter space. On the high $R_0$ -- low-$\alpha$ side of the line, the behaviour is well represented by the non-behavioural limit, in which it is not rational to significantly modify one's behaviour. On the low $R_0$ -- high infection cost side, it is rational to strongly modify one's behaviour. The lines describing the crossover are given by the critical costs $\alpha_s^\star$ for the transition in the excess cases, see \cref{eq:excess_cases_crossover}, and/or $\alpha_i^\star$ for the transition in the infection peak, see \cref{eq:peaks_crossover}.
The parameter values used for some of the curves in \cref{fig:Fig1,fig:Fig2} are marked by analogously coloured dots.
}
	   \label{fig:alpha_star}
\end{figure}

\section{Results}

The higher the infection cost $\alpha$, the stronger is the incentive to reduce social activity and hence $k$, see \cref{fig:Fig1}. The stronger the reduction in $k$, the more slowly the epidemic progresses, the lower the peak infection levels are, and the higher the total number of cases $1- s_f$ becomes.

In what follows, we analyse the epidemic using two key quantities, the excess cases $\eps$ and the peak of the epidemic $\max(i)$. 
For $t_f \to \infty$, herd immunity is always reached. The final number of susceptibles then always satisfies 
$s_f \leq 1/R_0$, with $1/R_0$ the minimum number of cases for which herd immunity is guaranteed. 
The cases in excess of this threshold are defined as
\begin{align}
\eps = 1/R_0 - s_f
\label{eq:excess_cases_def}
\end{align}
We will calculate $\eps$ and $\max(i)$ in two limiting cases: 
(1) The \textit{Non-behavioural limit} in which there is no perceived infection cost $\alpha = 0$.  In this case there is no reason to modify one's behaviour, $k = R_0$, see purple lines in \cref{fig:Fig2,fig:Fig1}. 
(2) The \textit{high-infection-cost asymptote} in which infection costs are very high, $\alpha/R_0^2 \gg 1$.
By matching these solutions we will obtain crossover costs between these scaling results.

\subsection*{Non-behavioural limit}
For this edge case only, the analytic solution was known previously \cite{Kermack1927,Miller2012,Harko2014,Miller2017,Kroger2020}.
We recover it in our notation as follows. 
Since $\alpha = 0$, \cref{eq:beautiful} is solved by
\begin{align}
	s(r) = s_0e^{-R_0 (r-r_0)}
	\label{eq:solution_nonbehavioural}
\end{align}
Its limit $s_f = e^{-R_0 (1-s_f-r_0)}$
yields 
\begin{align}
s_f = - W(- s_0 R_0 e^{R_0(r_0-1)})/R_0
\label{eq:sf_nonbehavioural}
\end{align}
with the product logarithm $W$. 
Hence,
\begin{align}
\eps = (1 + W(-s_0R_0e^{R_0(r_0-1)}))/R_0.
\label{eq:excess_cases_nonbehavioural}
\end{align}
The peak of the epidemic $\hat i = \max(i) = i(\hat t)$ occurs at the time $\hat t$ for which $\dot i(\hat t) = 0$ and thus
 $s(\hat t) = 1/R_0$, see \cref{dimlessgeneral}. Inserting this and $r = 1-s-i$ into \cref{eq:solution_nonbehavioural}, we obtain 
\begin{gather}
 \hat i = \max(i) =  1 - r_0 - (1+\ln(s_0R_0))/R_0
 \label{eq:peak_nonbehavioural}
\end{gather}

\subsection*{High-infection-cost asymptote} 
The final number of cases $s_f$ can be calculated in the limit of large $\alpha \gg R_0^2$, where $s_f = 1/R_0 - \eps$ with $\eps$ small and assuming that $s_0 > 1/R_0$. We obtain
\begin{align}
	\eps = 2R_0^2/\alpha
	\label{eq:excess_cases_high_alpha}
\end{align}
from an expansion of \cref{eq:analytic_solution} in both $1/\alpha$ and $\eps$ small and by matching order by order. This result is satisfied well, see \cref{fig:excess_cases}a. 
For the peak height, we obtain in the same limit, see \cref{sec:peak_height} for the calculation,
\begin{align}
	\hat i  = \max(i) = 2 R_0 (R_0 - 1)/\alpha
	\label{eq:peak_high_alpha}
\end{align}
which is also satisfied well, see \cref{fig:infection_peak}b.

\subsection*{Scaling and phase diagram} 
Observing in  \cref{fig:excess_cases}a that the excess cases are roughly constant at low $\alpha$ and therefore well described by the non-behavioural limit, we obtain a crossover cost $\alpha_s^\star$ at which the non-behavioural and high infection cost asymptotes of \cref{eq:excess_cases_nonbehavioural,eq:excess_cases_high_alpha}, respectively, match
\begin{align}
	\alpha_s^\star = 2R_0^3/\left(1+(W(-s_0R_0e^{R_0(r_0-1)})\right)
	\label{eq:excess_cases_crossover}
\end{align}
For the infection peak, we similarly obtain a crossover cost $\alpha_i^\star$ from matching \cref{eq:peak_nonbehavioural,eq:peak_high_alpha}
\begin{align}
\alpha_i^\star = 2 R_0^2 (R_0 - 1)/\left(R_0(1 - r_0) - 1 - \ln(s_0R_0)\right)
	\label{eq:peaks_crossover}
\end{align}
These crossover values and the non-behavioural limits for $\eps$ and $\max(i)$ can be used to achieve complete collapse of $\eps$ and $\max(i)$ onto master curves, see \cref{fig:excess_cases}b and \cref{fig:infection_peak}d, respectively.

Both crossover values, $\alpha_i^\star$ and $\alpha_s^\star$, determine different aspects of the ``phase diagram'' of social distancing, see \cref{fig:alpha_star}.
The crossover $\alpha_i^\star$ for the infection peak describes a behavioural transition in the most intuitive signal of an epidemic. 
The infection peak also corresponds to the most restrictive value of social distancing, see \cref{eq:k_vs_i}. For $\alpha < \alpha_i^\star$ social distancing is extremely weak, see e.g. for $\alpha=50$ in \cref{fig:Fig1}a (For $R_0 = 4$, $\alpha_i^\star \approx 59$ and $\alpha_s^\star \approx 139$).
Social distancing is ultimately aimed at reducing excess cases. For $\alpha_i^\star \leq \alpha \leq \alpha_s^\star$ there is social distancing, but still only on a relatively short time frame, see the data for $\alpha=100$ in \cref{fig:Fig1}a.
It starts to visibly affect the peak of the epidemic but not its duration, \cref{fig:Fig1}c, and has a very limited effect on the total of cases, \cref{fig:Fig1}b. This can be a viewed as the consequence of low gearing between the drop in infectivity and excess cases. 
Only for $\alpha > \alpha_s^\star$ is there considerable social distancing for an extended time, which then achieves a significant reduction in excess cases.

\begin{figure}[tbp]
  \centering
    \includegraphics[width=\columnwidth]{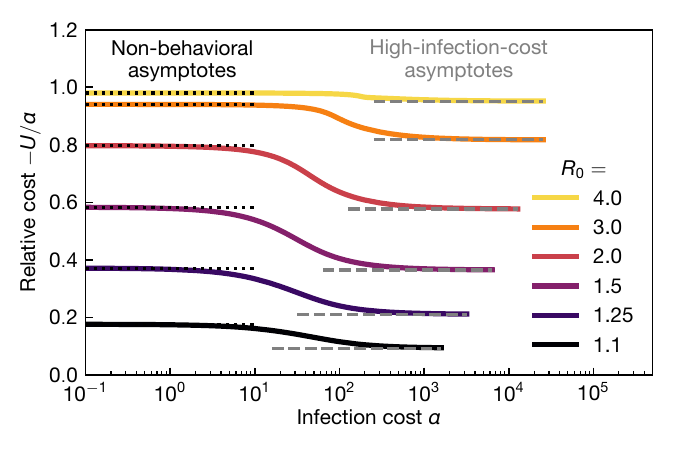}
  \caption{
	\textbf{Cost of the epidemic.} 
	Total epidemic cost relative to the cost of an infection, $-U/\alpha$, as a function of infection cost $\alpha$ under equilibrium social distancing. The corresponding non-behavioural, \cref{eq:utility_lowAlpha}, and high-infection-cost asymptotes, \cref{eq:utility_highAlpha} are indicated by dotted and dashed lines, respectively.
	}
  \label{fig:utility}
\end{figure}

\subsection*{Utility}

The utility, \cref{eq:U_with_salvage}, evaluated at the equilibrium behaviour can be directly calculated using the analytic solution
\begin{align}
U
&= -\alpha\left[r_f - r_0 + \frac{s_f}{2}\left(R_0(r_f-r_0) +  \ln(s_f/s_0)\right)\right]
\label{eq:utility_soln}
\end{align}
noting that $s_f$ and $r_f = 1-s_f$ depend on $\alpha$ and $R_0$. The total infection cost is given by $-\alpha(r_f-r_0)$ with the remainder being the total social distancing cost. Especially for intermediary $R_0$ and high infection costs $\alpha$, equilibrium behaviour strongly reduces the total epidemic cost, see \cref{fig:utility}. 
Again, we investigate the two limiting cases: For low $\alpha$, we obtain with 
\cref{eq:sf_nonbehavioural}
\begin{align}
U 
&= -\alpha\left[1 + \frac{1}{R_0}W(- s_0 R_0 e^{R_0(r_0-1)}) - r_0\right]
\label{eq:utility_lowAlpha}
\end{align}
For high $\alpha$, i.e. $\alpha/R_0^2 \gg 1$, with 
$s_f = \frac{1}{R_0} - \eps = \frac{1}{R_0} - \frac{2 R_0^2}{\alpha} \approx 1/R_0$, we obtain
\begin{align}
U
&= -\alpha\left[\frac{3}{2}(1- r_0)   - \frac{3 +  \ln(R_0s_0)}{2R_0}\right]
\label{eq:utility_highAlpha}
\end{align}

\section{Discussion and conclusion}

In summary, we have identified an analytic solution
for the Nash equilibrium behaviour for social distancing during an epidemic. 
We leveraged this solution to obtain
(1) a simple relationship between the strength of rational social distancing and the current number of cases, \cref{eq:k_vs_i}; 
(2) scaling results for the total number of cases, \cref{eq:excess_cases_high_alpha}, and 
the infection peak, \cref{eq:peak_high_alpha} 
which only depend on the basic reproduction number and the cost of contracting the disease; 
(3) characteristic infection costs, \cref{eq:excess_cases_crossover,eq:peaks_crossover}, that divide regimes of strong and weak social distancing and depend only on the basic reproduction number of the disease; 
(4) a closed form expression for the value of the utility, \cref{eq:utility_soln}.
These four results represent a remarkable simplification of a complex optimisation problem. 

We believe our work to be useful to policy makers because it yields a simple, albeit idealised, classification of the impact of self-organised social distancing during epidemics and thus can serve as guide for policy.
Given the basic reproduction number of a given disease $R_0$ and its estimated cost of infection $\alpha$, we show that one can either expect negligible social distancing from the population, when the infection cost is below a characteristic cost, or substantial social distancing when it is above.

There is an ongoing debate about the degree to which behaviour of individuals is truly rational, as we (and others) assume. In this context  our most significant result is that the rational decision making process {seems to be intuitively accessible to most members of the population}: {\bf rational social distancing is proportional to the infection cost and to the current number of cases}.
It is remarkable that the rational response we derive can be condensed into such a simple heuristic, understandable to a typical member of the population. While it may indeed be a challenge for such individuals to derive our results for themselves, a policymaker could communicate this simple heuristic, to be adopted by the population in order to assist them in targeting truly rational behaviour. It is not unrealistic to expect  this advice to influence the population decision making, especially given that it can be shown to be in each individual's self interest. In this sense the present work may itself help to ``bootstrap'' such rational behaviour.

While rational behaviour is not the mathematically optimal solution that maximises utility, as would be accessible under arbitrarily precise government control, it is relatively close to it. Rational behaviour also has the advantage of being stable, in the sense that it suppresses the detrimental behaviour of freeloaders, who are worse off if they deviate from the Nash equilibrium behaviour. The fact that rational behaviour is so desirable means that new tools that enable policymakers to help individuals target rational behaviour, like the ones we provide here, may be extremely valuable.

\section{Methods}

\subsection{Vaccination salvage term}
\label{sec:salvage_term}
A perfect vaccine applied to the whole population at time $t_f$ corresponds to immediately moving the susceptible fraction of the population into the recovered compartment, $s(t>t_f) = 0$ and $\psi_s(t>t_f) = 0$. Eqs. (\ref{dimlessgeneral}) reduce to 
\begin{align}
\dot i&= - i
\end{align}
The remaining infectious recover exponentially, with $i(t_f) = i_f$,
\begin{align}
	i(t > t_f) = i_f \exp[-(t-t_f)]
\end{align} 
Analogously for the individual probabilities, with $\psi_i(t_f) = \psi_{i,f}$,
\begin{align}
	\psi_i(t > t_f) = \psi_{i,f} \exp[-(t-t_f)]
	\label{eq:psi_i_asymptote}
\end{align}
Since nobody can get freshly infected, the population selects pre-epidemic behaviour, $\kappa(t>t_f)= R_0$. The contribution to the utility $U_f$ that arises from the recovery process after $t_f$ can be written in analogy to  \cref{Uda}
\begin{align*}
U_f&=\int_{t_f}^{\infty} f^{-t}\left[-\alpha\> \psi_i(t)-\betaGhost(\kd(t)-R_0)^2\right] dt
\end{align*}
This can be integrated to yield
\begin{align}
U_f = -f^{-t_f}\alpha\ \frac{\psi_{i,f}}{1+\ln f}
	\label{eq:individual_salvage_term_faccination}
\end{align}

\subsection{High-infection cost asymptote for the infection peak height}
\label{sec:peak_height}
In the large $\alpha$ limit $\alpha/R_0^2 \gg 1$ we have $s_f \approx 1/R_0$ from \cref{eq:excess_cases_def}, hence
\begin{align}
b = \alpha s_f /2 \approx \alpha/(2R_0)	
\label{eq:b_high_alpha}
\end{align}
large according to $b \gg R_0$. The infection peak $\dot i = 0$ occurs at $\hat i = \max(i)$ where \cref{dimlessgeneral} yields $0 = ks - 1$. 
Using \cref{eq:k_vs_i} we have
\begin{align}
	1 = ks =  s (R_0 - \hat ib) \Rightarrow s = \frac{1}{R_0 - \hat ib}
\end{align}
with the sum rule,
\begin{gather}
	\hat i = 1- r -s  = 1-r - \frac{1}{R_0 - \hat ib} \cr
\Rightarrow	(R_0 - \hat i b)\hat i = (1-r)(R_0 - \hat ib)-1
\end{gather}
This yields a quadratic equation for $\hat i$ with physical root
\begin{align}
\hat i 
=\ &\frac{R_0 + b(1-r)}{2b} \cr 
&- \frac{\sqrt{(R_0 + b(1-r))^2 + 4b(1-(1-r)R_0)}}{2b} \cr
	\approx\ &\frac{(1-r)}{2} \left[\frac{4b((1-r)R_0-1)}{2(R_0 + b(1-r))^2}\right] 
\end{align}
For large $\alpha$  the infection peak occurs early in the epidemic, when $r = 1 - s - i\ll 1$, 
e.g. see \cref{fig:Fig2} for $\alpha=400$. Using \cref{eq:b_high_alpha} and recalling also that $b \gg R_0$ we find
\begin{align}
\hat i	&\approx 
\frac{2R_0(R_0 - 1)}{\alpha}
\end{align}

\acknowledgments
We would like to dedicate this work to the memory of Prof. George Rowlands, who passed away in early 2021 and who was involved in many of the early discussions leading to this work.  
We thank Paul François, Shuhei Horiguchi, Tetsuya J. Kobayashi, and Takehiro Tottori for helpful discussions. 

This work was supported by the Grants-in-Aid for Scientific Research (JSPS KAKENHI) under Grants No. 20H00129 (RY), 20H05619 (RY), 22H04841 (SKS), 22K14012 (SKS), 23H04508 (JJM), and the JSPS Core-to-Core Program “Advanced core-to-core network for the physics of self-organizing active matter” JPJSCCA20230002 (all of us). MST acknowledges the generous support of visiting fellowships from JSPS Fellowship, ID L19547, the Leverhulme Trust, Ref. IAF-2019-019, and the kind hospitality of the Yamamoto group.
The funders had no role in study design, data collection and analysis, decision to publish, or preparation of the manuscript. 
The authors declare that there are no conflicts of interest.

Author contributions: All authors designed the research. SKS and MST performed the research. JJM assisted with code development and numerical methods. All authors wrote the paper.

\bibliography{paper-rational-policy}

\begin{thebibliography}{75}%
\makeatletter
\providecommand \@ifxundefined [1]{%
 \@ifx{#1\undefined}
}%
\providecommand \@ifnum [1]{%
 \ifnum #1\expandafter \@firstoftwo
 \else \expandafter \@secondoftwo
 \fi
}%
\providecommand \@ifx [1]{%
 \ifx #1\expandafter \@firstoftwo
 \else \expandafter \@secondoftwo
 \fi
}%
\providecommand \natexlab [1]{#1}%
\providecommand \enquote  [1]{``#1''}%
\providecommand \bibnamefont  [1]{#1}%
\providecommand \bibfnamefont [1]{#1}%
\providecommand \citenamefont [1]{#1}%
\providecommand \href@noop [0]{\@secondoftwo}%
\providecommand \href [0]{\begingroup \@sanitize@url \@href}%
\providecommand \@href[1]{\@@startlink{#1}\@@href}%
\providecommand \@@href[1]{\endgroup#1\@@endlink}%
\providecommand \@sanitize@url [0]{\catcode `\\12\catcode `\$12\catcode
  `\&12\catcode `\#12\catcode `\^12\catcode `\_12\catcode `\%12\relax}%
\providecommand \@@startlink[1]{}%
\providecommand \@@endlink[0]{}%
\providecommand \url  [0]{\begingroup\@sanitize@url \@url }%
\providecommand \@url [1]{\endgroup\@href {#1}{\urlprefix }}%
\providecommand \urlprefix  [0]{URL }%
\providecommand \Eprint [0]{\href }%
\providecommand \doibase [0]{https://doi.org/}%
\providecommand \selectlanguage [0]{\@gobble}%
\providecommand \bibinfo  [0]{\@secondoftwo}%
\providecommand \bibfield  [0]{\@secondoftwo}%
\providecommand \translation [1]{[#1]}%
\providecommand \BibitemOpen [0]{}%
\providecommand \bibitemStop [0]{}%
\providecommand \bibitemNoStop [0]{.\EOS\space}%
\providecommand \EOS [0]{\spacefactor3000\relax}%
\providecommand \BibitemShut  [1]{\csname bibitem#1\endcsname}%
\let\auto@bib@innerbib\@empty
%</preamble>
\bibitem [{\citenamefont {Reluga}(2010)}]{Reluga2010}%
  \BibitemOpen
  \bibfield  {author} {\bibinfo {author} {\bibfnamefont {T.~C.}\ \bibnamefont
  {Reluga}},\ }\bibfield  {title} {\bibinfo {title} {{Game Theory of Social
  Distancing in Response to an Epidemic}},\ }\href
  {https://doi.org/10.1371/journal.pcbi.1000793} {\bibfield  {journal}
  {\bibinfo  {journal} {PLoS Comput. Biol.}\ }\textbf {\bibinfo {volume} {6}},\
  \bibinfo {pages} {e1000793} (\bibinfo {year} {2010})}\BibitemShut {NoStop}%
\bibitem [{\citenamefont {Fenichel}\ \emph {et~al.}(2011)\citenamefont
  {Fenichel}, \citenamefont {Castillo-Chavez}, \citenamefont {Ceddia},
  \citenamefont {Chowell}, \citenamefont {Parra}, \citenamefont {Hickling},
  \citenamefont {Holloway}, \citenamefont {Horan}, \citenamefont {Morin},
  \citenamefont {Perrings}, \citenamefont {Springborn}, \citenamefont
  {Velazquez},\ and\ \citenamefont {Villalobos}}]{Fenichel2011}%
  \BibitemOpen
  \bibfield  {author} {\bibinfo {author} {\bibfnamefont {E.~P.}\ \bibnamefont
  {Fenichel}}, \bibinfo {author} {\bibfnamefont {C.}~\bibnamefont
  {Castillo-Chavez}}, \bibinfo {author} {\bibfnamefont {M.~G.}\ \bibnamefont
  {Ceddia}}, \bibinfo {author} {\bibfnamefont {G.}~\bibnamefont {Chowell}},
  \bibinfo {author} {\bibfnamefont {P.~A.~G.}\ \bibnamefont {Parra}}, \bibinfo
  {author} {\bibfnamefont {G.~J.}\ \bibnamefont {Hickling}}, \bibinfo {author}
  {\bibfnamefont {G.}~\bibnamefont {Holloway}}, \bibinfo {author}
  {\bibfnamefont {R.}~\bibnamefont {Horan}}, \bibinfo {author} {\bibfnamefont
  {B.}~\bibnamefont {Morin}}, \bibinfo {author} {\bibfnamefont
  {C.}~\bibnamefont {Perrings}}, \bibinfo {author} {\bibfnamefont
  {M.}~\bibnamefont {Springborn}}, \bibinfo {author} {\bibfnamefont
  {L.}~\bibnamefont {Velazquez}},\ and\ \bibinfo {author} {\bibfnamefont
  {C.}~\bibnamefont {Villalobos}},\ }\bibfield  {title} {\bibinfo {title}
  {{Adaptive human behavior in epidemiological models}},\ }\href
  {https://doi.org/10.1073/pnas.1011250108} {\bibfield  {journal} {\bibinfo
  {journal} {Proc. Natl. Acad. Sci.}\ }\textbf {\bibinfo {volume} {108}},\
  \bibinfo {pages} {6306} (\bibinfo {year} {2011})}\BibitemShut {NoStop}%
\bibitem [{\citenamefont {Wang}\ \emph {et~al.}(2016)\citenamefont {Wang},
  \citenamefont {Bauch}, \citenamefont {Bhattacharyya}, \citenamefont
  {D'Onofrio}, \citenamefont {Manfredi}, \citenamefont {Perc}, \citenamefont
  {Perra}, \citenamefont {Salath{\'{e}}},\ and\ \citenamefont
  {Zhao}}]{Wang2016}%
  \BibitemOpen
  \bibfield  {author} {\bibinfo {author} {\bibfnamefont {Z.}~\bibnamefont
  {Wang}}, \bibinfo {author} {\bibfnamefont {C.~T.}\ \bibnamefont {Bauch}},
  \bibinfo {author} {\bibfnamefont {S.}~\bibnamefont {Bhattacharyya}}, \bibinfo
  {author} {\bibfnamefont {A.}~\bibnamefont {D'Onofrio}}, \bibinfo {author}
  {\bibfnamefont {P.}~\bibnamefont {Manfredi}}, \bibinfo {author}
  {\bibfnamefont {M.}~\bibnamefont {Perc}}, \bibinfo {author} {\bibfnamefont
  {N.}~\bibnamefont {Perra}}, \bibinfo {author} {\bibfnamefont
  {M.}~\bibnamefont {Salath{\'{e}}}},\ and\ \bibinfo {author} {\bibfnamefont
  {D.}~\bibnamefont {Zhao}},\ }\bibfield  {title} {\bibinfo {title}
  {{Statistical physics of vaccination}},\ }\href
  {https://doi.org/10.1016/j.physrep.2016.10.006} {\bibfield  {journal}
  {\bibinfo  {journal} {Physics Reports}\ }\textbf {\bibinfo {volume} {664}},\
  \bibinfo {pages} {1} (\bibinfo {year} {2016})},\ \Eprint
  {https://arxiv.org/abs/1608.09010} {arXiv:1608.09010} \BibitemShut {NoStop}%
\bibitem [{\citenamefont {Chang}\ \emph {et~al.}(2020)\citenamefont {Chang},
  \citenamefont {Piraveenan}, \citenamefont {Pattison},\ and\ \citenamefont
  {Prokopenko}}]{Chang2020}%
  \BibitemOpen
  \bibfield  {author} {\bibinfo {author} {\bibfnamefont {S.~L.}\ \bibnamefont
  {Chang}}, \bibinfo {author} {\bibfnamefont {M.}~\bibnamefont {Piraveenan}},
  \bibinfo {author} {\bibfnamefont {P.}~\bibnamefont {Pattison}},\ and\
  \bibinfo {author} {\bibfnamefont {M.}~\bibnamefont {Prokopenko}},\ }\bibfield
   {title} {\bibinfo {title} {{Game theoretic modelling of infectious disease
  dynamics and intervention methods: a review}},\ }\href
  {https://doi.org/10.1080/17513758.2020.1720322} {\bibfield  {journal}
  {\bibinfo  {journal} {Journal of Biological Dynamics}\ }\textbf {\bibinfo
  {volume} {14}},\ \bibinfo {pages} {57} (\bibinfo {year} {2020})},\ \Eprint
  {https://arxiv.org/abs/1901.04143} {arXiv:1901.04143} \BibitemShut {NoStop}%
\bibitem [{\citenamefont {Verelst}\ \emph {et~al.}(2016)\citenamefont
  {Verelst}, \citenamefont {Willem},\ and\ \citenamefont
  {Beutels}}]{Verelst2016}%
  \BibitemOpen
  \bibfield  {author} {\bibinfo {author} {\bibfnamefont {F.}~\bibnamefont
  {Verelst}}, \bibinfo {author} {\bibfnamefont {L.}~\bibnamefont {Willem}},\
  and\ \bibinfo {author} {\bibfnamefont {P.}~\bibnamefont {Beutels}},\
  }\bibfield  {title} {\bibinfo {title} {{Behavioural change models for
  infectious disease transmission: A systematic review (2010-2015)}},\
  }\bibfield  {journal} {\bibinfo  {journal} {Journal of the Royal Society
  Interface}\ }\textbf {\bibinfo {volume} {13}},\ \href
  {https://doi.org/10.1098/rsif.2016.0820} {10.1098/rsif.2016.0820} (\bibinfo
  {year} {2016})\BibitemShut {NoStop}%
\bibitem [{\citenamefont {Bhattacharyya}\ and\ \citenamefont
  {Reluga}(2019)}]{Bhattacharyya2019}%
  \BibitemOpen
  \bibfield  {author} {\bibinfo {author} {\bibfnamefont {S.}~\bibnamefont
  {Bhattacharyya}}\ and\ \bibinfo {author} {\bibfnamefont {T.}~\bibnamefont
  {Reluga}},\ }\bibfield  {title} {\bibinfo {title} {{Game dynamic model of
  social distancing while cost of infection varies with epidemic burden}},\
  }\href {https://doi.org/10.1093/imamat/hxy047} {\bibfield  {journal}
  {\bibinfo  {journal} {IMA J. Appl. Math. (Institute Math. Its Appl.}\
  }\textbf {\bibinfo {volume} {84}},\ \bibinfo {pages} {23} (\bibinfo {year}
  {2019})}\BibitemShut {NoStop}%
\bibitem [{\citenamefont {Makris}\ and\ \citenamefont
  {Toxvaerd}(2020)}]{Makris2020}%
  \BibitemOpen
  \bibfield  {author} {\bibinfo {author} {\bibfnamefont {M.}~\bibnamefont
  {Makris}}\ and\ \bibinfo {author} {\bibfnamefont {F.}~\bibnamefont
  {Toxvaerd}},\ }\href {https://doi.org/10.17863/CAM.62310} {\emph {\bibinfo
  {title} {Cambridge Working Papers in Economics}}},\ \bibinfo {type}
  {Cambridge Working Papers in Economics}\ \bibinfo {number} {2097}\ (\bibinfo
  {year} {2020})\BibitemShut {NoStop}%
\bibitem [{\citenamefont {Yan}\ \emph {et~al.}(2021)\citenamefont {Yan},
  \citenamefont {Malik}, \citenamefont {Bayham}, \citenamefont {Fenichel},
  \citenamefont {Couzens},\ and\ \citenamefont {Omer}}]{Yan2021}%
  \BibitemOpen
  \bibfield  {author} {\bibinfo {author} {\bibfnamefont {Y.}~\bibnamefont
  {Yan}}, \bibinfo {author} {\bibfnamefont {A.~A.}\ \bibnamefont {Malik}},
  \bibinfo {author} {\bibfnamefont {J.}~\bibnamefont {Bayham}}, \bibinfo
  {author} {\bibfnamefont {E.~P.}\ \bibnamefont {Fenichel}}, \bibinfo {author}
  {\bibfnamefont {C.}~\bibnamefont {Couzens}},\ and\ \bibinfo {author}
  {\bibfnamefont {S.~B.}\ \bibnamefont {Omer}},\ }\bibfield  {title} {\bibinfo
  {title} {{Measuring voluntary and policy-induced social distancing behavior
  during the COVID-19 pandemic}},\ }\href
  {https://doi.org/10.1073/pnas.2008814118} {\bibfield  {journal} {\bibinfo
  {journal} {Proceedings of the National Academy of Sciences of the United
  States of America}\ }\textbf {\bibinfo {volume} {118}},\ \bibinfo {pages} {1}
  (\bibinfo {year} {2021})}\BibitemShut {NoStop}%
\bibitem [{\citenamefont {Reluga}\ and\ \citenamefont
  {Galvani}(2011)}]{Reluga2011}%
  \BibitemOpen
  \bibfield  {author} {\bibinfo {author} {\bibfnamefont {T.~C.}\ \bibnamefont
  {Reluga}}\ and\ \bibinfo {author} {\bibfnamefont {A.~P.}\ \bibnamefont
  {Galvani}},\ }\bibfield  {title} {\bibinfo {title} {{A general approach for
  population games with application to vaccination}},\ }\href
  {https://doi.org/10.1016/j.mbs.2011.01.003} {\bibfield  {journal} {\bibinfo
  {journal} {Math. Biosci.}\ }\textbf {\bibinfo {volume} {230}},\ \bibinfo
  {pages} {67} (\bibinfo {year} {2011})}\BibitemShut {NoStop}%
\bibitem [{\citenamefont {Lux}(2021)}]{Lux2021}%
  \BibitemOpen
  \bibfield  {author} {\bibinfo {author} {\bibfnamefont {T.}~\bibnamefont
  {Lux}},\ }\bibfield  {title} {\bibinfo {title} {{The social dynamics of
  COVID-19}},\ }\href {https://doi.org/10.1016/j.physa.2020.125710} {\bibfield
  {journal} {\bibinfo  {journal} {Physica A: Statistical Mechanics and its
  Applications}\ }\textbf {\bibinfo {volume} {567}},\ \bibinfo {pages} {125710}
  (\bibinfo {year} {2021})}\BibitemShut {NoStop}%
\bibitem [{\citenamefont {Kahneman}(2003)}]{kahneman2003maps}%
  \BibitemOpen
  \bibfield  {author} {\bibinfo {author} {\bibfnamefont {D.}~\bibnamefont
  {Kahneman}},\ }\bibfield  {title} {\bibinfo {title} {Maps of bounded
  rationality: Psychology for behavioral economics},\ }\href@noop {} {\bibfield
   {journal} {\bibinfo  {journal} {American economic review}\ }\textbf
  {\bibinfo {volume} {93}},\ \bibinfo {pages} {1449} (\bibinfo {year}
  {2003})}\BibitemShut {NoStop}%
\bibitem [{\citenamefont {McAdams}(2020)}]{McAdams2020}%
  \BibitemOpen
  \bibfield  {author} {\bibinfo {author} {\bibfnamefont {D.}~\bibnamefont
  {McAdams}},\ }\bibfield  {title} {\bibinfo {title} {{Nash SIR: An
  Economic-Epidemiological Model of Strategic Behavior During a Viral
  Epidemic}},\ }\bibfield  {journal} {\bibinfo  {journal} {Covid Economics}\
  }\href {https://doi.org/10.2139/ssrn.3593272} {10.2139/ssrn.3593272}
  (\bibinfo {year} {2020})\BibitemShut {NoStop}%
\bibitem [{\citenamefont {Eichenbaum}\ \emph {et~al.}(2021)\citenamefont
  {Eichenbaum}, \citenamefont {Rebelo},\ and\ \citenamefont
  {Trabandt}}]{Eichenbaum2021}%
  \BibitemOpen
  \bibfield  {author} {\bibinfo {author} {\bibfnamefont {M.~S.}\ \bibnamefont
  {Eichenbaum}}, \bibinfo {author} {\bibfnamefont {S.}~\bibnamefont {Rebelo}},\
  and\ \bibinfo {author} {\bibfnamefont {M.}~\bibnamefont {Trabandt}},\
  }\bibfield  {title} {\bibinfo {title} {{The Macroeconomics of Epidemics}},\
  }\href {https://doi.org/10.1093/rfs/hhab040} {\bibfield  {journal} {\bibinfo
  {journal} {Review of Financial Studies}\ }\textbf {\bibinfo {volume} {34}},\
  \bibinfo {pages} {5149} (\bibinfo {year} {2021})}\BibitemShut {NoStop}%
\bibitem [{\citenamefont {Rowthorn}\ and\ \citenamefont
  {Toxvaerd}(2020)}]{Rowthorn2020}%
  \BibitemOpen
  \bibfield  {author} {\bibinfo {author} {\bibfnamefont {R.}~\bibnamefont
  {Rowthorn}}\ and\ \bibinfo {author} {\bibfnamefont {F.}~\bibnamefont
  {Toxvaerd}},\ }\href@noop {} {\emph {\bibinfo {title} {{The optimal control
  of infectious diseases via prevention and treatment}}}},\ \bibinfo {type}
  {Cambridge Working Papers in Economics}\ \bibinfo {number} {2027}\ (\bibinfo
  {institution} {University of Cambridge},\ \bibinfo {year} {2020})\BibitemShut
  {NoStop}%
\bibitem [{\citenamefont {Toxvaerd}\ and\ \citenamefont
  {Rowthorn}(2020)}]{Toxvaerd2020}%
  \BibitemOpen
  \bibfield  {author} {\bibinfo {author} {\bibfnamefont {F.}~\bibnamefont
  {Toxvaerd}}\ and\ \bibinfo {author} {\bibfnamefont {R.}~\bibnamefont
  {Rowthorn}},\ }\href@noop {} {\emph {\bibinfo {title} {{On the management of
  population immunity}}}},\ \bibinfo {type} {Cambridge Working Papers in
  Economics}\ \bibinfo {number} {2080}\ (\bibinfo  {institution} {University of
  Cambridge},\ \bibinfo {year} {2020})\BibitemShut {NoStop}%
\bibitem [{\citenamefont {Li}\ \emph {et~al.}(2017)\citenamefont {Li},
  \citenamefont {Lindberg}, \citenamefont {Smith},\ and\ \citenamefont
  {Reluga}}]{Li2017}%
  \BibitemOpen
  \bibfield  {author} {\bibinfo {author} {\bibfnamefont {J.}~\bibnamefont
  {Li}}, \bibinfo {author} {\bibfnamefont {D.~V.}\ \bibnamefont {Lindberg}},
  \bibinfo {author} {\bibfnamefont {R.~A.}\ \bibnamefont {Smith}},\ and\
  \bibinfo {author} {\bibfnamefont {T.~C.}\ \bibnamefont {Reluga}},\ }\bibfield
   {title} {\bibinfo {title} {{Provisioning of Public Health Can Be Designed to
  Anticipate Public Policy Responses}},\ }\href
  {https://doi.org/10.1007/s11538-016-0231-8} {\bibfield  {journal} {\bibinfo
  {journal} {Bull. Math. Biol.}\ }\textbf {\bibinfo {volume} {79}},\ \bibinfo
  {pages} {163} (\bibinfo {year} {2017})}\BibitemShut {NoStop}%
\bibitem [{\citenamefont {Bethune}\ and\ \citenamefont
  {Korinek}(2020)}]{Bethune2020}%
  \BibitemOpen
  \bibfield  {author} {\bibinfo {author} {\bibfnamefont {Z.~A.}\ \bibnamefont
  {Bethune}}\ and\ \bibinfo {author} {\bibfnamefont {A.}~\bibnamefont
  {Korinek}},\ }\href@noop {} {\emph {\bibinfo {title} {{COVID-19 infection
  externalities: trading off lives vs. livelihoods}}}},\ \bibinfo {type}
  {Working Paper}\ \bibinfo {number} {27009}\ (\bibinfo  {institution}
  {National Bureau of Economic Research},\ \bibinfo {address} {Cambridge, MA},\
  \bibinfo {year} {2020})\BibitemShut {NoStop}%
\bibitem [{\citenamefont {Aurell}\ \emph {et~al.}(2022)\citenamefont {Aurell},
  \citenamefont {Carmona}, \citenamefont {Dayanikli},\ and\ \citenamefont
  {Lauri{\`{e}}re}}]{Aurell2022}%
  \BibitemOpen
  \bibfield  {author} {\bibinfo {author} {\bibfnamefont {A.}~\bibnamefont
  {Aurell}}, \bibinfo {author} {\bibfnamefont {R.}~\bibnamefont {Carmona}},
  \bibinfo {author} {\bibfnamefont {G.}~\bibnamefont {Dayanikli}},\ and\
  \bibinfo {author} {\bibfnamefont {M.}~\bibnamefont {Lauri{\`{e}}re}},\
  }\bibfield  {title} {\bibinfo {title} {{Optimal Incentives to Mitigate
  Epidemics: A Stackelberg Mean Field Game Approach}},\ }\href
  {https://doi.org/10.1137/20M1377862} {\bibfield  {journal} {\bibinfo
  {journal} {SIAM Journal on Control and Optimization}\ }\textbf {\bibinfo
  {volume} {60}},\ \bibinfo {pages} {S294} (\bibinfo {year} {2022})},\ \Eprint
  {https://arxiv.org/abs/2011.03105} {arXiv:2011.03105} \BibitemShut {NoStop}%
\bibitem [{\citenamefont {Schnyder}\ \emph
  {et~al.}(2023{\natexlab{a}})\citenamefont {Schnyder}, \citenamefont {Molina},
  \citenamefont {Yamamoto},\ and\ \citenamefont {Turner}}]{Schnyder2023b}%
  \BibitemOpen
  \bibfield  {author} {\bibinfo {author} {\bibfnamefont {S.~K.}\ \bibnamefont
  {Schnyder}}, \bibinfo {author} {\bibfnamefont {J.~J.}\ \bibnamefont
  {Molina}}, \bibinfo {author} {\bibfnamefont {R.}~\bibnamefont {Yamamoto}},\
  and\ \bibinfo {author} {\bibfnamefont {M.~S.}\ \bibnamefont {Turner}},\
  }\bibfield  {title} {\bibinfo {title} {{Rational social distancing policy
  during epidemics with limited healthcare capacity}},\ }\href
  {https://doi.org/10.1371/journal.pcbi.1011533} {\bibfield  {journal}
  {\bibinfo  {journal} {PLOS Computational Biology}\ }\textbf {\bibinfo
  {volume} {19}},\ \bibinfo {pages} {e1011533} (\bibinfo {year}
  {2023}{\natexlab{a}})},\ \Eprint {https://arxiv.org/abs/2205.00684}
  {arXiv:2205.00684} \BibitemShut {NoStop}%
\bibitem [{\citenamefont {Althouse}\ \emph {et~al.}(2010)\citenamefont
  {Althouse}, \citenamefont {Bergstrom},\ and\ \citenamefont
  {Bergstrom}}]{Althouse2010}%
  \BibitemOpen
  \bibfield  {author} {\bibinfo {author} {\bibfnamefont {B.~M.}\ \bibnamefont
  {Althouse}}, \bibinfo {author} {\bibfnamefont {T.~C.}\ \bibnamefont
  {Bergstrom}},\ and\ \bibinfo {author} {\bibfnamefont {C.~T.}\ \bibnamefont
  {Bergstrom}},\ }\bibfield  {title} {\bibinfo {title} {{A public choice
  framework for controlling transmissible and evolving diseases}},\ }\href
  {https://doi.org/10.1073/pnas.0906078107} {\bibfield  {journal} {\bibinfo
  {journal} {Proceedings of the National Academy of Sciences of the United
  States of America}\ }\textbf {\bibinfo {volume} {107}},\ \bibinfo {pages}
  {1696} (\bibinfo {year} {2010})}\BibitemShut {NoStop}%
\bibitem [{\citenamefont {Acemoglu}\ \emph {et~al.}(2020)\citenamefont
  {Acemoglu}, \citenamefont {Chernozhukov}, \citenamefont {Werning},\ and\
  \citenamefont {Whinston}}]{Acemoglu2020}%
  \BibitemOpen
  \bibfield  {author} {\bibinfo {author} {\bibfnamefont {D.}~\bibnamefont
  {Acemoglu}}, \bibinfo {author} {\bibfnamefont {V.}~\bibnamefont
  {Chernozhukov}}, \bibinfo {author} {\bibfnamefont {I.}~\bibnamefont
  {Werning}},\ and\ \bibinfo {author} {\bibfnamefont {M.~D.}\ \bibnamefont
  {Whinston}},\ }\href {https://doi.org/10.3386/w27102} {\emph {\bibinfo
  {title} {{Optimal targeted lockdowns in a multi-group SIR model}}}},\
  \bibinfo {type} {Working Paper}\ \bibinfo {number} {27102}\ (\bibinfo
  {institution} {National Bureau of Economic Research},\ \bibinfo {year}
  {2020})\BibitemShut {NoStop}%
\bibitem [{\citenamefont {Prem}\ \emph {et~al.}(2017)\citenamefont {Prem},
  \citenamefont {Cook},\ and\ \citenamefont {Jit}}]{Prem2017}%
  \BibitemOpen
  \bibfield  {author} {\bibinfo {author} {\bibfnamefont {K.}~\bibnamefont
  {Prem}}, \bibinfo {author} {\bibfnamefont {A.~R.}\ \bibnamefont {Cook}},\
  and\ \bibinfo {author} {\bibfnamefont {M.}~\bibnamefont {Jit}},\ }\bibfield
  {title} {\bibinfo {title} {{Projecting social contact matrices in 152
  countries using contact surveys and demographic data}},\ }\href
  {https://doi.org/10.1371/journal.pcbi.1005697} {\bibfield  {journal}
  {\bibinfo  {journal} {PLOS Computational Biology}\ }\textbf {\bibinfo
  {volume} {13}},\ \bibinfo {pages} {e1005697} (\bibinfo {year}
  {2017})}\BibitemShut {NoStop}%
\bibitem [{\citenamefont {Huang}\ \emph {et~al.}(2022)\citenamefont {Huang},
  \citenamefont {Crump}, \citenamefont {Brown}, \citenamefont {Spencer},
  \citenamefont {Miaka}, \citenamefont {Shampa}, \citenamefont {Keeling},\ and\
  \citenamefont {Rock}}]{Huang2022}%
  \BibitemOpen
  \bibfield  {author} {\bibinfo {author} {\bibfnamefont {C.~I.}\ \bibnamefont
  {Huang}}, \bibinfo {author} {\bibfnamefont {R.~E.}\ \bibnamefont {Crump}},
  \bibinfo {author} {\bibfnamefont {P.~E.}\ \bibnamefont {Brown}}, \bibinfo
  {author} {\bibfnamefont {S.~E.}\ \bibnamefont {Spencer}}, \bibinfo {author}
  {\bibfnamefont {E.~M.}\ \bibnamefont {Miaka}}, \bibinfo {author}
  {\bibfnamefont {C.}~\bibnamefont {Shampa}}, \bibinfo {author} {\bibfnamefont
  {M.~J.}\ \bibnamefont {Keeling}},\ and\ \bibinfo {author} {\bibfnamefont
  {K.~S.}\ \bibnamefont {Rock}},\ }\bibfield  {title} {\bibinfo {title}
  {{Identifying regions for enhanced control of gambiense sleeping sickness in
  the Democratic Republic of Congo}},\ }\href
  {https://doi.org/10.1038/s41467-022-29192-w} {\bibfield  {journal} {\bibinfo
  {journal} {Nature Communications}\ }\textbf {\bibinfo {volume} {13}},\
  \bibinfo {pages} {1} (\bibinfo {year} {2022})}\BibitemShut {NoStop}%
\bibitem [{\citenamefont {Tildesley}\ \emph {et~al.}(2022)\citenamefont
  {Tildesley}, \citenamefont {Vassall}, \citenamefont {Riley}, \citenamefont
  {Jit}, \citenamefont {Sandmann}, \citenamefont {Hill}, \citenamefont
  {Thompson}, \citenamefont {Atkins}, \citenamefont {Edmunds}, \citenamefont
  {Dyson},\ and\ \citenamefont {Keeling}}]{Tildesley2022}%
  \BibitemOpen
  \bibfield  {author} {\bibinfo {author} {\bibfnamefont {M.~J.}\ \bibnamefont
  {Tildesley}}, \bibinfo {author} {\bibfnamefont {A.}~\bibnamefont {Vassall}},
  \bibinfo {author} {\bibfnamefont {S.}~\bibnamefont {Riley}}, \bibinfo
  {author} {\bibfnamefont {M.}~\bibnamefont {Jit}}, \bibinfo {author}
  {\bibfnamefont {F.}~\bibnamefont {Sandmann}}, \bibinfo {author}
  {\bibfnamefont {E.~M.}\ \bibnamefont {Hill}}, \bibinfo {author}
  {\bibfnamefont {R.~N.}\ \bibnamefont {Thompson}}, \bibinfo {author}
  {\bibfnamefont {B.~D.}\ \bibnamefont {Atkins}}, \bibinfo {author}
  {\bibfnamefont {J.}~\bibnamefont {Edmunds}}, \bibinfo {author} {\bibfnamefont
  {L.}~\bibnamefont {Dyson}},\ and\ \bibinfo {author} {\bibfnamefont {M.~J.}\
  \bibnamefont {Keeling}},\ }\bibfield  {title} {\bibinfo {title} {{Optimal
  health and economic impact of non-pharmaceutical intervention measures prior
  and post vaccination in England: a mathematical modelling study}},\
  }\bibfield  {journal} {\bibinfo  {journal} {Royal Society Open Science}\
  }\textbf {\bibinfo {volume} {9}},\ \href
  {https://doi.org/10.1098/rsos.211746} {10.1098/rsos.211746} (\bibinfo {year}
  {2022})\BibitemShut {NoStop}%
\bibitem [{\citenamefont {Keeling}\ \emph {et~al.}(2022)\citenamefont
  {Keeling}, \citenamefont {Dyson}, \citenamefont {Tildesley}, \citenamefont
  {Hill},\ and\ \citenamefont {Moore}}]{Keeling2022}%
  \BibitemOpen
  \bibfield  {author} {\bibinfo {author} {\bibfnamefont {M.~J.}\ \bibnamefont
  {Keeling}}, \bibinfo {author} {\bibfnamefont {L.}~\bibnamefont {Dyson}},
  \bibinfo {author} {\bibfnamefont {M.~J.}\ \bibnamefont {Tildesley}}, \bibinfo
  {author} {\bibfnamefont {E.~M.}\ \bibnamefont {Hill}},\ and\ \bibinfo
  {author} {\bibfnamefont {S.}~\bibnamefont {Moore}},\ }\bibfield  {title}
  {\bibinfo {title} {{Comparison of the 2021 COVID-19 roadmap projections
  against public health data in England}},\ }\href
  {https://doi.org/10.1038/s41467-022-31991-0} {\bibfield  {journal} {\bibinfo
  {journal} {Nature communications}\ }\textbf {\bibinfo {volume} {13}},\
  \bibinfo {pages} {4924} (\bibinfo {year} {2022})}\BibitemShut {NoStop}%
\bibitem [{\citenamefont {He}\ \emph {et~al.}(2013)\citenamefont {He},
  \citenamefont {Dushoff}, \citenamefont {Day}, \citenamefont {Ma},\ and\
  \citenamefont {Earn}}]{He2013}%
  \BibitemOpen
  \bibfield  {author} {\bibinfo {author} {\bibfnamefont {D.}~\bibnamefont
  {He}}, \bibinfo {author} {\bibfnamefont {J.}~\bibnamefont {Dushoff}},
  \bibinfo {author} {\bibfnamefont {T.}~\bibnamefont {Day}}, \bibinfo {author}
  {\bibfnamefont {J.}~\bibnamefont {Ma}},\ and\ \bibinfo {author}
  {\bibfnamefont {D.~J.}\ \bibnamefont {Earn}},\ }\bibfield  {title} {\bibinfo
  {title} {{Inferring the causes of the three waves of the 1918 influenza
  pandemic in England and Wales}},\ }\bibfield  {journal} {\bibinfo  {journal}
  {Proceedings of the Royal Society B: Biological Sciences}\ }\textbf {\bibinfo
  {volume} {280}},\ \href {https://doi.org/10.1098/rspb.2013.1345}
  {10.1098/rspb.2013.1345} (\bibinfo {year} {2013})\BibitemShut {NoStop}%
\bibitem [{\citenamefont {Giannitsarou}\ \emph {et~al.}(2021)\citenamefont
  {Giannitsarou}, \citenamefont {Kissler},\ and\ \citenamefont
  {Toxvaerd}}]{Giannitsarou2021}%
  \BibitemOpen
  \bibfield  {author} {\bibinfo {author} {\bibfnamefont {C.}~\bibnamefont
  {Giannitsarou}}, \bibinfo {author} {\bibfnamefont {S.}~\bibnamefont
  {Kissler}},\ and\ \bibinfo {author} {\bibfnamefont {F.}~\bibnamefont
  {Toxvaerd}},\ }\bibfield  {title} {\bibinfo {title} {{Waning Immunity and the
  Second Wave: Some Projections for SARS-CoV-2}},\ }\href
  {https://doi.org/10.1257/aeri.20200343} {\bibfield  {journal} {\bibinfo
  {journal} {American Economic Review: Insights}\ }\textbf {\bibinfo {volume}
  {3}},\ \bibinfo {pages} {321} (\bibinfo {year} {2021})}\BibitemShut {NoStop}%
\bibitem [{\citenamefont {Schwarzendahl}\ \emph {et~al.}(2022)\citenamefont
  {Schwarzendahl}, \citenamefont {Grauer}, \citenamefont {Liebchen},\ and\
  \citenamefont {L{\"{o}}wen}}]{Schwarzendahl2022}%
  \BibitemOpen
  \bibfield  {author} {\bibinfo {author} {\bibfnamefont {F.~J.}\ \bibnamefont
  {Schwarzendahl}}, \bibinfo {author} {\bibfnamefont {J.}~\bibnamefont
  {Grauer}}, \bibinfo {author} {\bibfnamefont {B.}~\bibnamefont {Liebchen}},\
  and\ \bibinfo {author} {\bibfnamefont {H.}~\bibnamefont {L{\"{o}}wen}},\
  }\bibfield  {title} {\bibinfo {title} {{Mutation induced infection waves in
  diseases like COVID-19}},\ }\href
  {https://doi.org/10.1038/s41598-022-13137-w} {\bibfield  {journal} {\bibinfo
  {journal} {Scientific Reports}\ }\textbf {\bibinfo {volume} {12}},\ \bibinfo
  {pages} {1} (\bibinfo {year} {2022})}\BibitemShut {NoStop}%
\bibitem [{\citenamefont {Mossong}\ \emph {et~al.}(2008)\citenamefont
  {Mossong}, \citenamefont {Hens}, \citenamefont {Jit}, \citenamefont
  {Beutels}, \citenamefont {Auranen}, \citenamefont {Mikolajczyk},
  \citenamefont {Massari}, \citenamefont {Salmaso}, \citenamefont {Tomba},
  \citenamefont {Wallinga}, \citenamefont {Heijne}, \citenamefont
  {Sadkowska-Todys}, \citenamefont {Rosinska},\ and\ \citenamefont
  {Edmunds}}]{Mossong2008}%
  \BibitemOpen
  \bibfield  {author} {\bibinfo {author} {\bibfnamefont {J.}~\bibnamefont
  {Mossong}}, \bibinfo {author} {\bibfnamefont {N.}~\bibnamefont {Hens}},
  \bibinfo {author} {\bibfnamefont {M.}~\bibnamefont {Jit}}, \bibinfo {author}
  {\bibfnamefont {P.}~\bibnamefont {Beutels}}, \bibinfo {author} {\bibfnamefont
  {K.}~\bibnamefont {Auranen}}, \bibinfo {author} {\bibfnamefont
  {R.}~\bibnamefont {Mikolajczyk}}, \bibinfo {author} {\bibfnamefont
  {M.}~\bibnamefont {Massari}}, \bibinfo {author} {\bibfnamefont
  {S.}~\bibnamefont {Salmaso}}, \bibinfo {author} {\bibfnamefont {G.~S.}\
  \bibnamefont {Tomba}}, \bibinfo {author} {\bibfnamefont {J.}~\bibnamefont
  {Wallinga}}, \bibinfo {author} {\bibfnamefont {J.}~\bibnamefont {Heijne}},
  \bibinfo {author} {\bibfnamefont {M.}~\bibnamefont {Sadkowska-Todys}},
  \bibinfo {author} {\bibfnamefont {M.}~\bibnamefont {Rosinska}},\ and\
  \bibinfo {author} {\bibfnamefont {W.~J.}\ \bibnamefont {Edmunds}},\
  }\bibfield  {title} {\bibinfo {title} {{Social contacts and mixing patterns
  relevant to the spread of infectious diseases}},\ }\href
  {https://doi.org/10.1371/journal.pmed.0050074} {\bibfield  {journal}
  {\bibinfo  {journal} {PLoS Medicine}\ }\textbf {\bibinfo {volume} {5}},\
  \bibinfo {pages} {0381} (\bibinfo {year} {2008})}\BibitemShut {NoStop}%
\bibitem [{\citenamefont {Tildesley}\ \emph {et~al.}(2010)\citenamefont
  {Tildesley}, \citenamefont {House}, \citenamefont {Bruhn}, \citenamefont
  {Curry}, \citenamefont {O'Neil}, \citenamefont {Allpress}, \citenamefont
  {Smith},\ and\ \citenamefont {Keeling}}]{Tildesley2010}%
  \BibitemOpen
  \bibfield  {author} {\bibinfo {author} {\bibfnamefont {M.~J.}\ \bibnamefont
  {Tildesley}}, \bibinfo {author} {\bibfnamefont {T.~A.}\ \bibnamefont
  {House}}, \bibinfo {author} {\bibfnamefont {M.~C.}\ \bibnamefont {Bruhn}},
  \bibinfo {author} {\bibfnamefont {R.~J.}\ \bibnamefont {Curry}}, \bibinfo
  {author} {\bibfnamefont {M.}~\bibnamefont {O'Neil}}, \bibinfo {author}
  {\bibfnamefont {J.~L.}\ \bibnamefont {Allpress}}, \bibinfo {author}
  {\bibfnamefont {G.}~\bibnamefont {Smith}},\ and\ \bibinfo {author}
  {\bibfnamefont {M.~J.}\ \bibnamefont {Keeling}},\ }\bibfield  {title}
  {\bibinfo {title} {{Impact of spatial clustering on disease transmission and
  optimal control}},\ }\href {https://doi.org/10.1073/pnas.0909047107}
  {\bibfield  {journal} {\bibinfo  {journal} {Proceedings of the National
  Academy of Sciences of the United States of America}\ }\textbf {\bibinfo
  {volume} {107}},\ \bibinfo {pages} {1041} (\bibinfo {year}
  {2010})}\BibitemShut {NoStop}%
\bibitem [{\citenamefont {Sun}\ \emph {et~al.}(2021)\citenamefont {Sun},
  \citenamefont {Wang}, \citenamefont {Gao}, \citenamefont {Wang},
  \citenamefont {Luo}, \citenamefont {Ren}, \citenamefont {Zhan}, \citenamefont
  {Chen}, \citenamefont {Zhao}, \citenamefont {Huang}, \citenamefont {Sun},
  \citenamefont {Liu}, \citenamefont {Litvinova}, \citenamefont {Vespignani},
  \citenamefont {Ajelli}, \citenamefont {Viboud},\ and\ \citenamefont
  {Yu}}]{Sun2021}%
  \BibitemOpen
  \bibfield  {author} {\bibinfo {author} {\bibfnamefont {K.}~\bibnamefont
  {Sun}}, \bibinfo {author} {\bibfnamefont {W.}~\bibnamefont {Wang}}, \bibinfo
  {author} {\bibfnamefont {L.}~\bibnamefont {Gao}}, \bibinfo {author}
  {\bibfnamefont {Y.}~\bibnamefont {Wang}}, \bibinfo {author} {\bibfnamefont
  {K.}~\bibnamefont {Luo}}, \bibinfo {author} {\bibfnamefont {L.}~\bibnamefont
  {Ren}}, \bibinfo {author} {\bibfnamefont {Z.}~\bibnamefont {Zhan}}, \bibinfo
  {author} {\bibfnamefont {X.}~\bibnamefont {Chen}}, \bibinfo {author}
  {\bibfnamefont {S.}~\bibnamefont {Zhao}}, \bibinfo {author} {\bibfnamefont
  {Y.}~\bibnamefont {Huang}}, \bibinfo {author} {\bibfnamefont
  {Q.}~\bibnamefont {Sun}}, \bibinfo {author} {\bibfnamefont {Z.}~\bibnamefont
  {Liu}}, \bibinfo {author} {\bibfnamefont {M.}~\bibnamefont {Litvinova}},
  \bibinfo {author} {\bibfnamefont {A.}~\bibnamefont {Vespignani}}, \bibinfo
  {author} {\bibfnamefont {M.}~\bibnamefont {Ajelli}}, \bibinfo {author}
  {\bibfnamefont {C.}~\bibnamefont {Viboud}},\ and\ \bibinfo {author}
  {\bibfnamefont {H.}~\bibnamefont {Yu}},\ }\bibfield  {title} {\bibinfo
  {title} {{Transmission heterogeneities, kinetics, and controllability of
  SARS-CoV-2}},\ }\href {https://doi.org/10.1126/science.abe2424} {\bibfield
  {journal} {\bibinfo  {journal} {Science}\ }\textbf {\bibinfo {volume}
  {371}},\ \bibinfo {pages} {eabe2424} (\bibinfo {year} {2021})}\BibitemShut
  {NoStop}%
\bibitem [{\citenamefont {Hill}\ \emph {et~al.}(2023)\citenamefont {Hill},
  \citenamefont {Prosser}, \citenamefont {Brown}, \citenamefont {Ferguson},
  \citenamefont {Green}, \citenamefont {Kaler}, \citenamefont {Keeling},\ and\
  \citenamefont {Tildesley}}]{Hill2023}%
  \BibitemOpen
  \bibfield  {author} {\bibinfo {author} {\bibfnamefont {E.~M.}\ \bibnamefont
  {Hill}}, \bibinfo {author} {\bibfnamefont {N.~S.}\ \bibnamefont {Prosser}},
  \bibinfo {author} {\bibfnamefont {P.~E.}\ \bibnamefont {Brown}}, \bibinfo
  {author} {\bibfnamefont {E.}~\bibnamefont {Ferguson}}, \bibinfo {author}
  {\bibfnamefont {M.~J.}\ \bibnamefont {Green}}, \bibinfo {author}
  {\bibfnamefont {J.}~\bibnamefont {Kaler}}, \bibinfo {author} {\bibfnamefont
  {M.~J.}\ \bibnamefont {Keeling}},\ and\ \bibinfo {author} {\bibfnamefont
  {M.~J.}\ \bibnamefont {Tildesley}},\ }\bibfield  {title} {\bibinfo {title}
  {{Incorporating heterogeneity in farmer disease control behaviour into a
  livestock disease transmission model}},\ }\href
  {https://doi.org/10.1016/j.prevetmed.2023.106019} {\bibfield  {journal}
  {\bibinfo  {journal} {Preventive Veterinary Medicine}\ }\textbf {\bibinfo
  {volume} {219}},\ \bibinfo {pages} {106019} (\bibinfo {year}
  {2023})}\BibitemShut {NoStop}%
\bibitem [{\citenamefont {Chandrasekhar}\ \emph {et~al.}(2021)\citenamefont
  {Chandrasekhar}, \citenamefont {Goldsmith-Pinkham}, \citenamefont {Jackson},\
  and\ \citenamefont {Thau}}]{Chandrasekhar2021}%
  \BibitemOpen
  \bibfield  {author} {\bibinfo {author} {\bibfnamefont {A.~G.}\ \bibnamefont
  {Chandrasekhar}}, \bibinfo {author} {\bibfnamefont {P.}~\bibnamefont
  {Goldsmith-Pinkham}}, \bibinfo {author} {\bibfnamefont {M.~O.}\ \bibnamefont
  {Jackson}},\ and\ \bibinfo {author} {\bibfnamefont {S.}~\bibnamefont
  {Thau}},\ }\bibfield  {title} {\bibinfo {title} {{Interacting regional
  policies in containing a disease}},\ }\href
  {https://doi.org/10.1073/pnas.2021520118} {\bibfield  {journal} {\bibinfo
  {journal} {Proceedings of the National Academy of Sciences of the United
  States of America}\ }\textbf {\bibinfo {volume} {118}},\ \bibinfo {pages} {1}
  (\bibinfo {year} {2021})},\ \Eprint {https://arxiv.org/abs/2008.10745}
  {arXiv:2008.10745} \BibitemShut {NoStop}%
\bibitem [{\citenamefont {Holme}\ and\ \citenamefont
  {Saram{\"{a}}ki}(2012)}]{Holme2012}%
  \BibitemOpen
  \bibfield  {author} {\bibinfo {author} {\bibfnamefont {P.}~\bibnamefont
  {Holme}}\ and\ \bibinfo {author} {\bibfnamefont {J.}~\bibnamefont
  {Saram{\"{a}}ki}},\ }\bibfield  {title} {\bibinfo {title} {{Temporal
  networks}},\ }\href {https://doi.org/10.1016/j.physrep.2012.03.001}
  {\bibfield  {journal} {\bibinfo  {journal} {Physics Reports}\ }\textbf
  {\bibinfo {volume} {519}},\ \bibinfo {pages} {97} (\bibinfo {year} {2012})},\
  \Eprint {https://arxiv.org/abs/1108.1780} {arXiv:1108.1780} \BibitemShut
  {NoStop}%
\bibitem [{\citenamefont {Holme}\ and\ \citenamefont
  {Masuda}(2015)}]{Holme2015}%
  \BibitemOpen
  \bibfield  {author} {\bibinfo {author} {\bibfnamefont {P.}~\bibnamefont
  {Holme}}\ and\ \bibinfo {author} {\bibfnamefont {N.}~\bibnamefont {Masuda}},\
  }\bibfield  {title} {\bibinfo {title} {{The basic reproduction number as a
  predictor for epidemic outbreaks in temporal networks}},\ }\href
  {https://doi.org/10.1371/journal.pone.0120567} {\bibfield  {journal}
  {\bibinfo  {journal} {PLoS ONE}\ }\textbf {\bibinfo {volume} {10}},\ \bibinfo
  {pages} {1} (\bibinfo {year} {2015})},\ \Eprint
  {https://arxiv.org/abs/1407.6598} {arXiv:1407.6598} \BibitemShut {NoStop}%
\bibitem [{\citenamefont {Ferguson}\ \emph {et~al.}(2006)\citenamefont
  {Ferguson}, \citenamefont {Cummings}, \citenamefont {Fraser}, \citenamefont
  {Cajka}, \citenamefont {Cooley},\ and\ \citenamefont {Burke}}]{Ferguson2006}%
  \BibitemOpen
  \bibfield  {author} {\bibinfo {author} {\bibfnamefont {N.~M.}\ \bibnamefont
  {Ferguson}}, \bibinfo {author} {\bibfnamefont {D.~A.~T.}\ \bibnamefont
  {Cummings}}, \bibinfo {author} {\bibfnamefont {C.}~\bibnamefont {Fraser}},
  \bibinfo {author} {\bibfnamefont {J.~C.}\ \bibnamefont {Cajka}}, \bibinfo
  {author} {\bibfnamefont {P.~C.}\ \bibnamefont {Cooley}},\ and\ \bibinfo
  {author} {\bibfnamefont {D.~S.}\ \bibnamefont {Burke}},\ }\bibfield  {title}
  {\bibinfo {title} {{Strategies for mitigating an influenza pandemic}},\
  }\href {https://doi.org/10.1038/nature04795} {\bibfield  {journal} {\bibinfo
  {journal} {Nature}\ }\textbf {\bibinfo {volume} {442}},\ \bibinfo {pages}
  {448} (\bibinfo {year} {2006})}\BibitemShut {NoStop}%
\bibitem [{\citenamefont {Tanimoto}(2018)}]{Tanimoto2018}%
  \BibitemOpen
  \bibfield  {author} {\bibinfo {author} {\bibfnamefont {J.}~\bibnamefont
  {Tanimoto}},\ }\href {https://doi.org/10.1007/978-981-13-2769-8_4} {\emph
  {\bibinfo {title} {{Social Dilemma Analysis of the Spread of Infectious
  Disease}}}}\ (\bibinfo {year} {2018})\ pp.\ \bibinfo {pages}
  {155--216}\BibitemShut {NoStop}%
\bibitem [{\citenamefont {Mellacher}(2020)}]{Mellacher2020}%
  \BibitemOpen
  \bibfield  {author} {\bibinfo {author} {\bibfnamefont {P.}~\bibnamefont
  {Mellacher}},\ }\href@noop {} {\emph {\bibinfo {title} {{COVID-Town: An
  Integrated Economic-Epidemiological Agent-Based Model}}}},\ \bibinfo {type}
  {GSC discussion papers}\ \bibinfo {number} {23}\ (\bibinfo  {institution}
  {Graz Schumpeter Centre},\ \bibinfo {year} {2020})\BibitemShut {NoStop}%
\bibitem [{\citenamefont {Grauer}\ \emph {et~al.}(2020)\citenamefont {Grauer},
  \citenamefont {L{\"{o}}wen},\ and\ \citenamefont {Liebchen}}]{Grauer2020}%
  \BibitemOpen
  \bibfield  {author} {\bibinfo {author} {\bibfnamefont {J.}~\bibnamefont
  {Grauer}}, \bibinfo {author} {\bibfnamefont {H.}~\bibnamefont
  {L{\"{o}}wen}},\ and\ \bibinfo {author} {\bibfnamefont {B.}~\bibnamefont
  {Liebchen}},\ }\bibfield  {title} {\bibinfo {title} {{Strategic
  spatiotemporal vaccine distribution increases the survival rate in an
  infectious disease like Covid-19}},\ }\href
  {https://doi.org/10.1038/s41598-020-78447-3} {\bibfield  {journal} {\bibinfo
  {journal} {Scientific Reports}\ }\textbf {\bibinfo {volume} {10}},\ \bibinfo
  {pages} {1} (\bibinfo {year} {2020})},\ \Eprint
  {https://arxiv.org/abs/2005.04056} {arXiv:2005.04056} \BibitemShut {NoStop}%
\bibitem [{\citenamefont {Yong}\ and\ \citenamefont
  {Zhou}(1999)}]{yong1999stochastic}%
  \BibitemOpen
  \bibfield  {author} {\bibinfo {author} {\bibfnamefont {J.}~\bibnamefont
  {Yong}}\ and\ \bibinfo {author} {\bibfnamefont {X.~Y.}\ \bibnamefont
  {Zhou}},\ }\href@noop {} {\emph {\bibinfo {title} {Stochastic controls:
  Hamiltonian systems and HJB equations}}},\ Vol.~\bibinfo {volume} {43}\
  (\bibinfo  {publisher} {Springer Science \& Business Media},\ \bibinfo {year}
  {1999})\BibitemShut {NoStop}%
\bibitem [{\citenamefont {Lorch}\ \emph {et~al.}(2018)\citenamefont {Lorch},
  \citenamefont {De}, \citenamefont {Bhatt}, \citenamefont {Trouleau},
  \citenamefont {Upadhyay},\ and\ \citenamefont {Gomez-Rodriguez}}]{Lorch2018}%
  \BibitemOpen
  \bibfield  {author} {\bibinfo {author} {\bibfnamefont {L.}~\bibnamefont
  {Lorch}}, \bibinfo {author} {\bibfnamefont {A.}~\bibnamefont {De}}, \bibinfo
  {author} {\bibfnamefont {S.}~\bibnamefont {Bhatt}}, \bibinfo {author}
  {\bibfnamefont {W.}~\bibnamefont {Trouleau}}, \bibinfo {author}
  {\bibfnamefont {U.}~\bibnamefont {Upadhyay}},\ and\ \bibinfo {author}
  {\bibfnamefont {M.}~\bibnamefont {Gomez-Rodriguez}},\ }\bibfield  {title}
  {\bibinfo {title} {{Stochastic Optimal Control of Epidemic Processes in
  Networks}},\ }\href {http://arxiv.org/abs/1810.13043} {\  (\bibinfo {year}
  {2018})},\ \Eprint {https://arxiv.org/abs/1810.13043} {arXiv:1810.13043}
  \BibitemShut {NoStop}%
\bibitem [{\citenamefont {Tottori}\ and\ \citenamefont
  {Kobayashi}(2022)}]{Tottori2022}%
  \BibitemOpen
  \bibfield  {author} {\bibinfo {author} {\bibfnamefont {T.}~\bibnamefont
  {Tottori}}\ and\ \bibinfo {author} {\bibfnamefont {T.~J.}\ \bibnamefont
  {Kobayashi}},\ }\bibfield  {title} {\bibinfo {title} {{Memory-Limited
  Partially Observable Stochastic Control and Its Mean-Field Control
  Approach}},\ }\href {https://doi.org/10.3390/e24111599} {\bibfield  {journal}
  {\bibinfo  {journal} {Entropy}\ }\textbf {\bibinfo {volume} {24}},\ \bibinfo
  {pages} {1} (\bibinfo {year} {2022})},\ \Eprint
  {https://arxiv.org/abs/2203.10682} {arXiv:2203.10682} \BibitemShut {NoStop}%
\bibitem [{\citenamefont {Tottori}\ and\ \citenamefont
  {Kobayashi}(2023{\natexlab{a}})}]{Tottori2023b}%
  \BibitemOpen
  \bibfield  {author} {\bibinfo {author} {\bibfnamefont {T.}~\bibnamefont
  {Tottori}}\ and\ \bibinfo {author} {\bibfnamefont {T.~J.}\ \bibnamefont
  {Kobayashi}},\ }\bibfield  {title} {\bibinfo {title} {{Forward-Backward Sweep
  Method for the System of HJB-FP Equations in Memory-Limited Partially
  Observable Stochastic Control}},\ }\bibfield  {journal} {\bibinfo  {journal}
  {Entropy}\ }\textbf {\bibinfo {volume} {25}},\ \href
  {https://doi.org/10.3390/e25020208} {10.3390/e25020208} (\bibinfo {year}
  {2023}{\natexlab{a}})\BibitemShut {NoStop}%
\bibitem [{\citenamefont {Tottori}\ and\ \citenamefont
  {Kobayashi}(2023{\natexlab{b}})}]{Tottori2023}%
  \BibitemOpen
  \bibfield  {author} {\bibinfo {author} {\bibfnamefont {T.}~\bibnamefont
  {Tottori}}\ and\ \bibinfo {author} {\bibfnamefont {T.~J.}\ \bibnamefont
  {Kobayashi}},\ }\bibfield  {title} {\bibinfo {title} {{Decentralized
  Stochastic Control with Finite-Dimensional Memories: A Memory Limitation
  Approach}},\ }\href {https://doi.org/10.3390/e25050791} {\bibfield  {journal}
  {\bibinfo  {journal} {Entropy}\ }\textbf {\bibinfo {volume} {25}},\ \bibinfo
  {pages} {1} (\bibinfo {year} {2023}{\natexlab{b}})}\BibitemShut {NoStop}%
\bibitem [{\citenamefont {Barnett}\ \emph {et~al.}(2023)\citenamefont
  {Barnett}, \citenamefont {Buchak},\ and\ \citenamefont
  {Yannelis}}]{Barnett2023}%
  \BibitemOpen
  \bibfield  {author} {\bibinfo {author} {\bibfnamefont {M.}~\bibnamefont
  {Barnett}}, \bibinfo {author} {\bibfnamefont {G.}~\bibnamefont {Buchak}},\
  and\ \bibinfo {author} {\bibfnamefont {C.}~\bibnamefont {Yannelis}},\
  }\bibfield  {title} {\bibinfo {title} {{Epidemic responses under
  uncertainty}},\ }\href {https://doi.org/10.1073/pnas.2208111120} {\bibfield
  {journal} {\bibinfo  {journal} {Proceedings of the National Academy of
  Sciences of the United States of America}\ }\textbf {\bibinfo {volume}
  {120}},\ \bibinfo {pages} {1} (\bibinfo {year} {2023})}\BibitemShut {NoStop}%
\bibitem [{\citenamefont {Shea}\ \emph {et~al.}(2023)\citenamefont {Shea},
  \citenamefont {Borchering}, \citenamefont {Probert}, \citenamefont
  {Howerton}, \citenamefont {Bogich}, \citenamefont {Li}, \citenamefont {van
  Panhuis}, \citenamefont {Viboud}, \citenamefont {Agu{\'{a}}s}, \citenamefont
  {Belov}, \citenamefont {Bhargava}, \citenamefont {Cavany}, \citenamefont
  {Chang}, \citenamefont {Chen}, \citenamefont {Chen}, \citenamefont {Chen},
  \citenamefont {Chen}, \citenamefont {Childs}, \citenamefont {Chow},
  \citenamefont {Crooker}, \citenamefont {{Del Valle}}, \citenamefont
  {Espa{\~{n}}a}, \citenamefont {Fairchild}, \citenamefont {Gerkin},
  \citenamefont {Germann}, \citenamefont {Gu}, \citenamefont {Guan},
  \citenamefont {Guo}, \citenamefont {Hart}, \citenamefont {Hladish},
  \citenamefont {Hupert}, \citenamefont {Janies}, \citenamefont {Kerr},
  \citenamefont {Klein}, \citenamefont {Klein}, \citenamefont {Lin},
  \citenamefont {Manore}, \citenamefont {Meyers}, \citenamefont {Mittler},
  \citenamefont {Mu}, \citenamefont {N{\'{u}}{\~{n}}ez}, \citenamefont
  {Oidtman}, \citenamefont {Pasco}, \citenamefont {{Pastore y Piontti}},
  \citenamefont {Paul}, \citenamefont {Pearson}, \citenamefont {Perdomo},
  \citenamefont {Perkins}, \citenamefont {Pierce}, \citenamefont {Pillai},
  \citenamefont {Rael}, \citenamefont {Rosenfeld}, \citenamefont {Ross},
  \citenamefont {Spencer}, \citenamefont {Stoltzfus}, \citenamefont {Toh},
  \citenamefont {Vattikuti}, \citenamefont {Vespignani}, \citenamefont {Wang},
  \citenamefont {White}, \citenamefont {Xu}, \citenamefont {Yang},
  \citenamefont {Yogurtcu}, \citenamefont {Zhang}, \citenamefont {Zhao},
  \citenamefont {Zou}, \citenamefont {Ferrari}, \citenamefont {Pannell},
  \citenamefont {Tildesley}, \citenamefont {Seifarth}, \citenamefont {Johnson},
  \citenamefont {Biggerstaff}, \citenamefont {Johansson}, \citenamefont
  {Slayton}, \citenamefont {Levander}, \citenamefont {Stazer}, \citenamefont
  {Kerr},\ and\ \citenamefont {Runge}}]{Shea2023}%
  \BibitemOpen
  \bibfield  {author} {\bibinfo {author} {\bibfnamefont {K.}~\bibnamefont
  {Shea}}, \bibinfo {author} {\bibfnamefont {R.~K.}\ \bibnamefont
  {Borchering}}, \bibinfo {author} {\bibfnamefont {W.~J.~M.}\ \bibnamefont
  {Probert}}, \bibinfo {author} {\bibfnamefont {E.}~\bibnamefont {Howerton}},
  \bibinfo {author} {\bibfnamefont {T.~L.}\ \bibnamefont {Bogich}}, \bibinfo
  {author} {\bibfnamefont {S.-L.}\ \bibnamefont {Li}}, \bibinfo {author}
  {\bibfnamefont {W.~G.}\ \bibnamefont {van Panhuis}}, \bibinfo {author}
  {\bibfnamefont {C.}~\bibnamefont {Viboud}}, \bibinfo {author} {\bibfnamefont
  {R.}~\bibnamefont {Agu{\'{a}}s}}, \bibinfo {author} {\bibfnamefont {A.~A.}\
  \bibnamefont {Belov}}, \bibinfo {author} {\bibfnamefont {S.~H.}\ \bibnamefont
  {Bhargava}}, \bibinfo {author} {\bibfnamefont {S.~M.}\ \bibnamefont
  {Cavany}}, \bibinfo {author} {\bibfnamefont {J.~C.}\ \bibnamefont {Chang}},
  \bibinfo {author} {\bibfnamefont {C.}~\bibnamefont {Chen}}, \bibinfo {author}
  {\bibfnamefont {J.}~\bibnamefont {Chen}}, \bibinfo {author} {\bibfnamefont
  {S.}~\bibnamefont {Chen}}, \bibinfo {author} {\bibfnamefont {Y.}~\bibnamefont
  {Chen}}, \bibinfo {author} {\bibfnamefont {L.~M.}\ \bibnamefont {Childs}},
  \bibinfo {author} {\bibfnamefont {C.~C.}\ \bibnamefont {Chow}}, \bibinfo
  {author} {\bibfnamefont {I.}~\bibnamefont {Crooker}}, \bibinfo {author}
  {\bibfnamefont {S.~Y.}\ \bibnamefont {{Del Valle}}}, \bibinfo {author}
  {\bibfnamefont {G.}~\bibnamefont {Espa{\~{n}}a}}, \bibinfo {author}
  {\bibfnamefont {G.}~\bibnamefont {Fairchild}}, \bibinfo {author}
  {\bibfnamefont {R.~C.}\ \bibnamefont {Gerkin}}, \bibinfo {author}
  {\bibfnamefont {T.~C.}\ \bibnamefont {Germann}}, \bibinfo {author}
  {\bibfnamefont {Q.}~\bibnamefont {Gu}}, \bibinfo {author} {\bibfnamefont
  {X.}~\bibnamefont {Guan}}, \bibinfo {author} {\bibfnamefont {L.}~\bibnamefont
  {Guo}}, \bibinfo {author} {\bibfnamefont {G.~R.}\ \bibnamefont {Hart}},
  \bibinfo {author} {\bibfnamefont {T.~J.}\ \bibnamefont {Hladish}}, \bibinfo
  {author} {\bibfnamefont {N.}~\bibnamefont {Hupert}}, \bibinfo {author}
  {\bibfnamefont {D.}~\bibnamefont {Janies}}, \bibinfo {author} {\bibfnamefont
  {C.~C.}\ \bibnamefont {Kerr}}, \bibinfo {author} {\bibfnamefont {D.~J.}\
  \bibnamefont {Klein}}, \bibinfo {author} {\bibfnamefont {E.~Y.}\ \bibnamefont
  {Klein}}, \bibinfo {author} {\bibfnamefont {G.}~\bibnamefont {Lin}}, \bibinfo
  {author} {\bibfnamefont {C.}~\bibnamefont {Manore}}, \bibinfo {author}
  {\bibfnamefont {L.~A.}\ \bibnamefont {Meyers}}, \bibinfo {author}
  {\bibfnamefont {J.~E.}\ \bibnamefont {Mittler}}, \bibinfo {author}
  {\bibfnamefont {K.}~\bibnamefont {Mu}}, \bibinfo {author} {\bibfnamefont
  {R.~C.}\ \bibnamefont {N{\'{u}}{\~{n}}ez}}, \bibinfo {author} {\bibfnamefont
  {R.~J.}\ \bibnamefont {Oidtman}}, \bibinfo {author} {\bibfnamefont
  {R.}~\bibnamefont {Pasco}}, \bibinfo {author} {\bibfnamefont
  {A.}~\bibnamefont {{Pastore y Piontti}}}, \bibinfo {author} {\bibfnamefont
  {R.}~\bibnamefont {Paul}}, \bibinfo {author} {\bibfnamefont {C.~A.~B.}\
  \bibnamefont {Pearson}}, \bibinfo {author} {\bibfnamefont {D.~R.}\
  \bibnamefont {Perdomo}}, \bibinfo {author} {\bibfnamefont {T.~A.}\
  \bibnamefont {Perkins}}, \bibinfo {author} {\bibfnamefont {K.}~\bibnamefont
  {Pierce}}, \bibinfo {author} {\bibfnamefont {A.~N.}\ \bibnamefont {Pillai}},
  \bibinfo {author} {\bibfnamefont {R.~C.}\ \bibnamefont {Rael}}, \bibinfo
  {author} {\bibfnamefont {K.}~\bibnamefont {Rosenfeld}}, \bibinfo {author}
  {\bibfnamefont {C.~W.}\ \bibnamefont {Ross}}, \bibinfo {author}
  {\bibfnamefont {J.~A.}\ \bibnamefont {Spencer}}, \bibinfo {author}
  {\bibfnamefont {A.~B.}\ \bibnamefont {Stoltzfus}}, \bibinfo {author}
  {\bibfnamefont {K.~B.}\ \bibnamefont {Toh}}, \bibinfo {author} {\bibfnamefont
  {S.}~\bibnamefont {Vattikuti}}, \bibinfo {author} {\bibfnamefont
  {A.}~\bibnamefont {Vespignani}}, \bibinfo {author} {\bibfnamefont
  {L.}~\bibnamefont {Wang}}, \bibinfo {author} {\bibfnamefont {L.~J.}\
  \bibnamefont {White}}, \bibinfo {author} {\bibfnamefont {P.}~\bibnamefont
  {Xu}}, \bibinfo {author} {\bibfnamefont {Y.}~\bibnamefont {Yang}}, \bibinfo
  {author} {\bibfnamefont {O.~N.}\ \bibnamefont {Yogurtcu}}, \bibinfo {author}
  {\bibfnamefont {W.}~\bibnamefont {Zhang}}, \bibinfo {author} {\bibfnamefont
  {Y.}~\bibnamefont {Zhao}}, \bibinfo {author} {\bibfnamefont {D.}~\bibnamefont
  {Zou}}, \bibinfo {author} {\bibfnamefont {M.~J.}\ \bibnamefont {Ferrari}},
  \bibinfo {author} {\bibfnamefont {D.}~\bibnamefont {Pannell}}, \bibinfo
  {author} {\bibfnamefont {M.~J.}\ \bibnamefont {Tildesley}}, \bibinfo {author}
  {\bibfnamefont {J.}~\bibnamefont {Seifarth}}, \bibinfo {author}
  {\bibfnamefont {E.}~\bibnamefont {Johnson}}, \bibinfo {author} {\bibfnamefont
  {M.}~\bibnamefont {Biggerstaff}}, \bibinfo {author} {\bibfnamefont {M.~A.}\
  \bibnamefont {Johansson}}, \bibinfo {author} {\bibfnamefont {R.~B.}\
  \bibnamefont {Slayton}}, \bibinfo {author} {\bibfnamefont {J.~D.}\
  \bibnamefont {Levander}}, \bibinfo {author} {\bibfnamefont {J.}~\bibnamefont
  {Stazer}}, \bibinfo {author} {\bibfnamefont {J.}~\bibnamefont {Kerr}},\ and\
  \bibinfo {author} {\bibfnamefont {M.~C.}\ \bibnamefont {Runge}},\ }\bibfield
  {title} {\bibinfo {title} {{Multiple models for outbreak decision support in
  the face of uncertainty}},\ }\href {https://doi.org/10.1073/pnas.2207537120}
  {\bibfield  {journal} {\bibinfo  {journal} {Proceedings of the National
  Academy of Sciences}\ }\textbf {\bibinfo {volume} {120}},\ \bibinfo {pages}
  {2017} (\bibinfo {year} {2023})}\BibitemShut {NoStop}%
\bibitem [{\citenamefont {Kantner}\ and\ \citenamefont
  {Koprucki}(2020)}]{Kantner2020}%
  \BibitemOpen
  \bibfield  {author} {\bibinfo {author} {\bibfnamefont {M.}~\bibnamefont
  {Kantner}}\ and\ \bibinfo {author} {\bibfnamefont {T.}~\bibnamefont
  {Koprucki}},\ }\bibfield  {title} {\bibinfo {title} {{Beyond just
  “flattening the curve”: Optimal control of epidemics with purely
  non-pharmaceutical interventions}},\ }\bibfield  {journal} {\bibinfo
  {journal} {Journal of Mathematics in Industry}\ }\textbf {\bibinfo {volume}
  {10}},\ \href {https://doi.org/10.1186/s13362-020-00091-3}
  {10.1186/s13362-020-00091-3} (\bibinfo {year} {2020}),\ \Eprint
  {https://arxiv.org/abs/2004.09471} {arXiv:2004.09471} \BibitemShut {NoStop}%
\bibitem [{\citenamefont {K{\"{o}}hler}\ \emph {et~al.}(2021)\citenamefont
  {K{\"{o}}hler}, \citenamefont {Schwenkel}, \citenamefont {Koch},
  \citenamefont {Berberich}, \citenamefont {Pauli},\ and\ \citenamefont
  {Allg{\"{o}}wer}}]{Kohler2021}%
  \BibitemOpen
  \bibfield  {author} {\bibinfo {author} {\bibfnamefont {J.}~\bibnamefont
  {K{\"{o}}hler}}, \bibinfo {author} {\bibfnamefont {L.}~\bibnamefont
  {Schwenkel}}, \bibinfo {author} {\bibfnamefont {A.}~\bibnamefont {Koch}},
  \bibinfo {author} {\bibfnamefont {J.}~\bibnamefont {Berberich}}, \bibinfo
  {author} {\bibfnamefont {P.}~\bibnamefont {Pauli}},\ and\ \bibinfo {author}
  {\bibfnamefont {F.}~\bibnamefont {Allg{\"{o}}wer}},\ }\bibfield  {title}
  {\bibinfo {title} {{Robust and optimal predictive control of the COVID-19
  outbreak}},\ }\href {https://doi.org/10.1016/j.arcontrol.2020.11.002}
  {\bibfield  {journal} {\bibinfo  {journal} {Annual Reviews in Control}\
  }\textbf {\bibinfo {volume} {51}},\ \bibinfo {pages} {525} (\bibinfo {year}
  {2021})},\ \Eprint {https://arxiv.org/abs/2005.03580} {arXiv:2005.03580}
  \BibitemShut {NoStop}%
\bibitem [{\citenamefont {Morris}\ \emph {et~al.}(2021)\citenamefont {Morris},
  \citenamefont {Rossine}, \citenamefont {Plotkin},\ and\ \citenamefont
  {Levin}}]{Morris2021}%
  \BibitemOpen
  \bibfield  {author} {\bibinfo {author} {\bibfnamefont {D.~H.}\ \bibnamefont
  {Morris}}, \bibinfo {author} {\bibfnamefont {F.~W.}\ \bibnamefont {Rossine}},
  \bibinfo {author} {\bibfnamefont {J.~B.}\ \bibnamefont {Plotkin}},\ and\
  \bibinfo {author} {\bibfnamefont {S.~A.}\ \bibnamefont {Levin}},\ }\bibfield
  {title} {\bibinfo {title} {{Optimal, near-optimal, and robust epidemic
  control}},\ }\href {https://doi.org/10.1038/s42005-021-00570-y} {\bibfield
  {journal} {\bibinfo  {journal} {Communications Physics}\ }\textbf {\bibinfo
  {volume} {4}},\ \bibinfo {pages} {1} (\bibinfo {year} {2021})},\ \Eprint
  {https://arxiv.org/abs/2004.02209} {arXiv:2004.02209} \BibitemShut {NoStop}%
\bibitem [{\citenamefont {Adhikari}\ \emph {et~al.}(2020)\citenamefont
  {Adhikari}, \citenamefont {Bolitho}, \citenamefont {Caballero}, \citenamefont
  {Cates}, \citenamefont {Dolezal}, \citenamefont {Ekeh}, \citenamefont
  {Guioth}, \citenamefont {Jack}, \citenamefont {Kappler}, \citenamefont
  {Kikuchi}, \citenamefont {Kobayashi}, \citenamefont {Li}, \citenamefont
  {Peterson}, \citenamefont {Pietzonka}, \citenamefont {Remez}, \citenamefont
  {Rohrbach}, \citenamefont {Singh},\ and\ \citenamefont
  {Turk}}]{Adhikari2020}%
  \BibitemOpen
  \bibfield  {author} {\bibinfo {author} {\bibfnamefont {R.}~\bibnamefont
  {Adhikari}}, \bibinfo {author} {\bibfnamefont {A.}~\bibnamefont {Bolitho}},
  \bibinfo {author} {\bibfnamefont {F.}~\bibnamefont {Caballero}}, \bibinfo
  {author} {\bibfnamefont {M.~E.}\ \bibnamefont {Cates}}, \bibinfo {author}
  {\bibfnamefont {J.}~\bibnamefont {Dolezal}}, \bibinfo {author} {\bibfnamefont
  {T.}~\bibnamefont {Ekeh}}, \bibinfo {author} {\bibfnamefont {J.}~\bibnamefont
  {Guioth}}, \bibinfo {author} {\bibfnamefont {R.~L.}\ \bibnamefont {Jack}},
  \bibinfo {author} {\bibfnamefont {J.}~\bibnamefont {Kappler}}, \bibinfo
  {author} {\bibfnamefont {L.}~\bibnamefont {Kikuchi}}, \bibinfo {author}
  {\bibfnamefont {H.}~\bibnamefont {Kobayashi}}, \bibinfo {author}
  {\bibfnamefont {Y.~I.}\ \bibnamefont {Li}}, \bibinfo {author} {\bibfnamefont
  {J.~D.}\ \bibnamefont {Peterson}}, \bibinfo {author} {\bibfnamefont
  {P.}~\bibnamefont {Pietzonka}}, \bibinfo {author} {\bibfnamefont
  {B.}~\bibnamefont {Remez}}, \bibinfo {author} {\bibfnamefont {P.~B.}\
  \bibnamefont {Rohrbach}}, \bibinfo {author} {\bibfnamefont {R.}~\bibnamefont
  {Singh}},\ and\ \bibinfo {author} {\bibfnamefont {G.}~\bibnamefont {Turk}},\
  }\bibfield  {title} {\bibinfo {title} {{Inference, prediction and
  optimization of non-pharmaceutical interventions using compartment models:
  the PyRoss library}},\ }\href {http://arxiv.org/abs/2005.09625} {\  (\bibinfo
  {year} {2020})},\ \Eprint {https://arxiv.org/abs/2005.09625}
  {arXiv:2005.09625} \BibitemShut {NoStop}%
\bibitem [{\citenamefont {Pietzonka}\ \emph {et~al.}(2021)\citenamefont
  {Pietzonka}, \citenamefont {Brorson}, \citenamefont {Bankes}, \citenamefont
  {Cates}, \citenamefont {Jack},\ and\ \citenamefont
  {Adhikari}}]{Pietzonka2021}%
  \BibitemOpen
  \bibfield  {author} {\bibinfo {author} {\bibfnamefont {P.}~\bibnamefont
  {Pietzonka}}, \bibinfo {author} {\bibfnamefont {E.}~\bibnamefont {Brorson}},
  \bibinfo {author} {\bibfnamefont {W.}~\bibnamefont {Bankes}}, \bibinfo
  {author} {\bibfnamefont {M.~E.}\ \bibnamefont {Cates}}, \bibinfo {author}
  {\bibfnamefont {R.~L.}\ \bibnamefont {Jack}},\ and\ \bibinfo {author}
  {\bibfnamefont {R.}~\bibnamefont {Adhikari}},\ }\bibfield  {title} {\bibinfo
  {title} {{Bayesian inference across multiple models suggests a strong
  increase in lethality of COVID-19 in late 2020 in the UK}},\ }\href
  {https://doi.org/10.1371/journal.pone.0258968} {\bibfield  {journal}
  {\bibinfo  {journal} {PLoS ONE}\ }\textbf {\bibinfo {volume} {16}},\ \bibinfo
  {pages} {1} (\bibinfo {year} {2021})}\BibitemShut {NoStop}%
\bibitem [{\citenamefont {Li}\ \emph {et~al.}(2021)\citenamefont {Li},
  \citenamefont {Turk}, \citenamefont {Rohrbach}, \citenamefont {Pietzonka},
  \citenamefont {Kappler}, \citenamefont {Singh}, \citenamefont {Dolezal},
  \citenamefont {Ekeh}, \citenamefont {Kikuchi}, \citenamefont {Peterson},
  \citenamefont {Bolitho}, \citenamefont {Kobayashi}, \citenamefont {Cates},
  \citenamefont {Adhikari},\ and\ \citenamefont {Jack}}]{Li2021}%
  \BibitemOpen
  \bibfield  {author} {\bibinfo {author} {\bibfnamefont {Y.~I.}\ \bibnamefont
  {Li}}, \bibinfo {author} {\bibfnamefont {G.}~\bibnamefont {Turk}}, \bibinfo
  {author} {\bibfnamefont {P.~B.}\ \bibnamefont {Rohrbach}}, \bibinfo {author}
  {\bibfnamefont {P.}~\bibnamefont {Pietzonka}}, \bibinfo {author}
  {\bibfnamefont {J.}~\bibnamefont {Kappler}}, \bibinfo {author} {\bibfnamefont
  {R.}~\bibnamefont {Singh}}, \bibinfo {author} {\bibfnamefont
  {J.}~\bibnamefont {Dolezal}}, \bibinfo {author} {\bibfnamefont
  {T.}~\bibnamefont {Ekeh}}, \bibinfo {author} {\bibfnamefont {L.}~\bibnamefont
  {Kikuchi}}, \bibinfo {author} {\bibfnamefont {J.~D.}\ \bibnamefont
  {Peterson}}, \bibinfo {author} {\bibfnamefont {A.}~\bibnamefont {Bolitho}},
  \bibinfo {author} {\bibfnamefont {H.}~\bibnamefont {Kobayashi}}, \bibinfo
  {author} {\bibfnamefont {M.~E.}\ \bibnamefont {Cates}}, \bibinfo {author}
  {\bibfnamefont {R.}~\bibnamefont {Adhikari}},\ and\ \bibinfo {author}
  {\bibfnamefont {R.~L.}\ \bibnamefont {Jack}},\ }\bibfield  {title} {\bibinfo
  {title} {{Efficient Bayesian inference of fully stochastic epidemiological
  models with applications to COVID-19}},\ }\href
  {https://doi.org/10.1098/rsos.211065} {\bibfield  {journal} {\bibinfo
  {journal} {Royal Society Open Science}\ }\textbf {\bibinfo {volume} {8}},\
  \bibinfo {pages} {211065} (\bibinfo {year} {2021})}\BibitemShut {NoStop}%
\bibitem [{\citenamefont {Molina}\ \emph {et~al.}(2022)\citenamefont {Molina},
  \citenamefont {Schnyder}, \citenamefont {Turner},\ and\ \citenamefont
  {Yamamoto}}]{Molina2022}%
  \BibitemOpen
  \bibfield  {author} {\bibinfo {author} {\bibfnamefont {J.~J.}\ \bibnamefont
  {Molina}}, \bibinfo {author} {\bibfnamefont {S.~K.}\ \bibnamefont
  {Schnyder}}, \bibinfo {author} {\bibfnamefont {M.~S.}\ \bibnamefont
  {Turner}},\ and\ \bibinfo {author} {\bibfnamefont {R.}~\bibnamefont
  {Yamamoto}},\ }\bibfield  {title} {\bibinfo {title} {{Nash Neural Networks :
  Inferring Utilities from Optimal Behaviour}},\ }\href
  {http://arxiv.org/abs/2203.13432} {\  (\bibinfo {year} {2022})},\ \Eprint
  {https://arxiv.org/abs/2203.13432} {arXiv:2203.13432} \BibitemShut {NoStop}%
\bibitem [{\citenamefont {Mellacher}(2023)}]{Mellacher2023}%
  \BibitemOpen
  \bibfield  {author} {\bibinfo {author} {\bibfnamefont {P.}~\bibnamefont
  {Mellacher}},\ }\bibfield  {title} {\bibinfo {title} {{The impact of corona
  populism: Empirical evidence from Austria and theory}},\ }\href
  {https://doi.org/10.1016/j.jebo.2023.02.021} {\bibfield  {journal} {\bibinfo
  {journal} {Journal of Economic Behavior \& Organization}\ }\textbf {\bibinfo
  {volume} {209}},\ \bibinfo {pages} {113} (\bibinfo {year}
  {2023})}\BibitemShut {NoStop}%
\bibitem [{\citenamefont {Kermack}\ and\ \citenamefont
  {McKendrick}(1927)}]{Kermack1927}%
  \BibitemOpen
  \bibfield  {author} {\bibinfo {author} {\bibfnamefont {W.~O.}\ \bibnamefont
  {Kermack}}\ and\ \bibinfo {author} {\bibfnamefont {A.}~\bibnamefont
  {McKendrick}},\ }\bibfield  {title} {\bibinfo {title} {{A contribution to the
  mathematical theory of epidemics}},\ }\href
  {https://doi.org/10.1098/rspa.1927.0118} {\bibfield  {journal} {\bibinfo
  {journal} {Proceedings of the Royal Society of London. Series A, Containing
  Papers of a Mathematical and Physical Character}\ }\textbf {\bibinfo {volume}
  {115}},\ \bibinfo {pages} {700} (\bibinfo {year} {1927})}\BibitemShut
  {NoStop}%
\bibitem [{\citenamefont {Miller}(2012)}]{Miller2012}%
  \BibitemOpen
  \bibfield  {author} {\bibinfo {author} {\bibfnamefont {J.~C.}\ \bibnamefont
  {Miller}},\ }\bibfield  {title} {\bibinfo {title} {{A Note on the Derivation
  of Epidemic Final Sizes}},\ }\href
  {https://doi.org/10.1007/s11538-012-9749-6} {\bibfield  {journal} {\bibinfo
  {journal} {Bulletin of Mathematical Biology}\ }\textbf {\bibinfo {volume}
  {74}},\ \bibinfo {pages} {2125} (\bibinfo {year} {2012})}\BibitemShut
  {NoStop}%
\bibitem [{\citenamefont {Harko}\ \emph {et~al.}(2014)\citenamefont {Harko},
  \citenamefont {Lobo},\ and\ \citenamefont {Mak}}]{Harko2014}%
  \BibitemOpen
  \bibfield  {author} {\bibinfo {author} {\bibfnamefont {T.}~\bibnamefont
  {Harko}}, \bibinfo {author} {\bibfnamefont {F.~S.}\ \bibnamefont {Lobo}},\
  and\ \bibinfo {author} {\bibfnamefont {M.~K.}\ \bibnamefont {Mak}},\
  }\bibfield  {title} {\bibinfo {title} {{Exact analytical solutions of the
  Susceptible-Infected-Recovered (SIR) epidemic model and of the SIR model with
  equal death and birth rates}},\ }\href
  {https://doi.org/10.1016/j.amc.2014.03.030} {\bibfield  {journal} {\bibinfo
  {journal} {Applied Mathematics and Computation}\ }\textbf {\bibinfo {volume}
  {236}},\ \bibinfo {pages} {184} (\bibinfo {year} {2014})},\ \Eprint
  {https://arxiv.org/abs/1403.2160} {arXiv:1403.2160} \BibitemShut {NoStop}%
\bibitem [{\citenamefont {Miller}(2017)}]{Miller2017}%
  \BibitemOpen
  \bibfield  {author} {\bibinfo {author} {\bibfnamefont {J.~C.}\ \bibnamefont
  {Miller}},\ }\bibfield  {title} {\bibinfo {title} {{Mathematical models of
  SIR disease spread with combined non-sexual and sexual transmission
  routes}},\ }\href {https://doi.org/10.1016/j.idm.2016.12.003} {\bibfield
  {journal} {\bibinfo  {journal} {Infectious Disease Modelling}\ }\textbf
  {\bibinfo {volume} {2}},\ \bibinfo {pages} {35} (\bibinfo {year} {2017})},\
  \Eprint {https://arxiv.org/abs/1609.08108} {1609.08108} \BibitemShut
  {NoStop}%
\bibitem [{\citenamefont {Kr{\"{o}}ger}\ and\ \citenamefont
  {Schlickeiser}(2020)}]{Kroger2020}%
  \BibitemOpen
  \bibfield  {author} {\bibinfo {author} {\bibfnamefont {M.}~\bibnamefont
  {Kr{\"{o}}ger}}\ and\ \bibinfo {author} {\bibfnamefont {R.}~\bibnamefont
  {Schlickeiser}},\ }\bibfield  {title} {\bibinfo {title} {{Analytical solution
  of the SIR-model for the temporal evolution of epidemics. Part A:
  time-independent reproduction factor}},\ }\bibfield  {journal} {\bibinfo
  {journal} {Journal of Physics A: Mathematical and Theoretical}\ }\textbf
  {\bibinfo {volume} {53}},\ \href {https://doi.org/10.1088/1751-8121/abc65d}
  {10.1088/1751-8121/abc65d} (\bibinfo {year} {2020})\BibitemShut {NoStop}%
\bibitem [{\citenamefont {Bauch}\ \emph {et~al.}(2003)\citenamefont {Bauch},
  \citenamefont {Galvani},\ and\ \citenamefont {Earn}}]{Bauch2003}%
  \BibitemOpen
  \bibfield  {author} {\bibinfo {author} {\bibfnamefont {C.~T.}\ \bibnamefont
  {Bauch}}, \bibinfo {author} {\bibfnamefont {A.~P.}\ \bibnamefont {Galvani}},\
  and\ \bibinfo {author} {\bibfnamefont {D.~J.}\ \bibnamefont {Earn}},\
  }\bibfield  {title} {\bibinfo {title} {{Group interest versus self-interest
  in smallpox vaccination policy}},\ }\href
  {https://doi.org/10.1073/pnas.1731324100} {\bibfield  {journal} {\bibinfo
  {journal} {Proceedings of the National Academy of Sciences of the United
  States of America}\ }\textbf {\bibinfo {volume} {100}},\ \bibinfo {pages}
  {10564} (\bibinfo {year} {2003})}\BibitemShut {NoStop}%
\bibitem [{\citenamefont {Bauch}\ and\ \citenamefont {Earn}(2004)}]{Bauch2004}%
  \BibitemOpen
  \bibfield  {author} {\bibinfo {author} {\bibfnamefont {C.~T.}\ \bibnamefont
  {Bauch}}\ and\ \bibinfo {author} {\bibfnamefont {D.~J.}\ \bibnamefont
  {Earn}},\ }\bibfield  {title} {\bibinfo {title} {{Vaccination and the theory
  of games}},\ }\href {https://doi.org/10.1073/pnas.0403823101} {\bibfield
  {journal} {\bibinfo  {journal} {Proceedings of the National Academy of
  Sciences of the United States of America}\ }\textbf {\bibinfo {volume}
  {101}},\ \bibinfo {pages} {13391} (\bibinfo {year} {2004})}\BibitemShut
  {NoStop}%
\bibitem [{\citenamefont {Reluga}\ \emph {et~al.}(2006)\citenamefont {Reluga},
  \citenamefont {Bauch},\ and\ \citenamefont {Galvani}}]{Reluga2006}%
  \BibitemOpen
  \bibfield  {author} {\bibinfo {author} {\bibfnamefont {T.~C.}\ \bibnamefont
  {Reluga}}, \bibinfo {author} {\bibfnamefont {C.~T.}\ \bibnamefont {Bauch}},\
  and\ \bibinfo {author} {\bibfnamefont {A.~P.}\ \bibnamefont {Galvani}},\
  }\bibfield  {title} {\bibinfo {title} {{Evolving public perceptions and
  stability in vaccine uptake}},\ }\href
  {https://doi.org/10.1016/j.mbs.2006.08.015} {\bibfield  {journal} {\bibinfo
  {journal} {Mathematical Biosciences}\ }\textbf {\bibinfo {volume} {204}},\
  \bibinfo {pages} {185} (\bibinfo {year} {2006})}\BibitemShut {NoStop}%
\bibitem [{\citenamefont {Tildesley}\ \emph {et~al.}(2006)\citenamefont
  {Tildesley}, \citenamefont {Savill}, \citenamefont {Shaw}, \citenamefont
  {Deardon}, \citenamefont {Brooks}, \citenamefont {Woolhouse}, \citenamefont
  {Grenfell},\ and\ \citenamefont {Keeling}}]{Tildesley2006}%
  \BibitemOpen
  \bibfield  {author} {\bibinfo {author} {\bibfnamefont {M.~J.}\ \bibnamefont
  {Tildesley}}, \bibinfo {author} {\bibfnamefont {N.~J.}\ \bibnamefont
  {Savill}}, \bibinfo {author} {\bibfnamefont {D.~J.}\ \bibnamefont {Shaw}},
  \bibinfo {author} {\bibfnamefont {R.}~\bibnamefont {Deardon}}, \bibinfo
  {author} {\bibfnamefont {S.~P.}\ \bibnamefont {Brooks}}, \bibinfo {author}
  {\bibfnamefont {M.~E.}\ \bibnamefont {Woolhouse}}, \bibinfo {author}
  {\bibfnamefont {B.~T.}\ \bibnamefont {Grenfell}},\ and\ \bibinfo {author}
  {\bibfnamefont {M.~J.}\ \bibnamefont {Keeling}},\ }\bibfield  {title}
  {\bibinfo {title} {{Optimal reactive vaccination strategies for a
  foot-and-mouth outbreak in the UK}},\ }\href
  {https://doi.org/10.1038/nature04324} {\bibfield  {journal} {\bibinfo
  {journal} {Nature}\ }\textbf {\bibinfo {volume} {440}},\ \bibinfo {pages}
  {83} (\bibinfo {year} {2006})}\BibitemShut {NoStop}%
\bibitem [{\citenamefont {Chen}\ and\ \citenamefont
  {Toxvaerd}(2014)}]{Chen2014}%
  \BibitemOpen
  \bibfield  {author} {\bibinfo {author} {\bibfnamefont {F.}~\bibnamefont
  {Chen}}\ and\ \bibinfo {author} {\bibfnamefont {F.}~\bibnamefont
  {Toxvaerd}},\ }\bibfield  {title} {\bibinfo {title} {{The economics of
  vaccination}},\ }\href {https://doi.org/10.1016/j.jtbi.2014.08.003}
  {\bibfield  {journal} {\bibinfo  {journal} {Journal of Theoretical Biology}\
  }\textbf {\bibinfo {volume} {363}},\ \bibinfo {pages} {105} (\bibinfo {year}
  {2014})}\BibitemShut {NoStop}%
\bibitem [{\citenamefont {Moore}\ \emph {et~al.}(2021)\citenamefont {Moore},
  \citenamefont {Hill}, \citenamefont {Dyson}, \citenamefont {Tildesley},\ and\
  \citenamefont {Keeling}}]{Moore2021b}%
  \BibitemOpen
  \bibfield  {author} {\bibinfo {author} {\bibfnamefont {S.}~\bibnamefont
  {Moore}}, \bibinfo {author} {\bibfnamefont {E.~M.}\ \bibnamefont {Hill}},
  \bibinfo {author} {\bibfnamefont {L.}~\bibnamefont {Dyson}}, \bibinfo
  {author} {\bibfnamefont {M.~J.}\ \bibnamefont {Tildesley}},\ and\ \bibinfo
  {author} {\bibfnamefont {M.~J.}\ \bibnamefont {Keeling}},\ }\bibfield
  {title} {\bibinfo {title} {{Modelling optimal vaccination strategy for
  SARS-CoV-2 in the UK}},\ }\href
  {https://doi.org/10.1371/journal.pcbi.1008849} {\bibfield  {journal}
  {\bibinfo  {journal} {PLoS Computational Biology}\ }\textbf {\bibinfo
  {volume} {17}},\ \bibinfo {pages} {1} (\bibinfo {year} {2021})}\BibitemShut
  {NoStop}%
\bibitem [{\citenamefont {Moore}\ \emph {et~al.}(2022)\citenamefont {Moore},
  \citenamefont {Hill}, \citenamefont {Dyson}, \citenamefont {Tildesley},\ and\
  \citenamefont {Keeling}}]{Moore2022}%
  \BibitemOpen
  \bibfield  {author} {\bibinfo {author} {\bibfnamefont {S.}~\bibnamefont
  {Moore}}, \bibinfo {author} {\bibfnamefont {E.~M.}\ \bibnamefont {Hill}},
  \bibinfo {author} {\bibfnamefont {L.}~\bibnamefont {Dyson}}, \bibinfo
  {author} {\bibfnamefont {M.~J.}\ \bibnamefont {Tildesley}},\ and\ \bibinfo
  {author} {\bibfnamefont {M.~J.}\ \bibnamefont {Keeling}},\ }\bibfield
  {title} {\bibinfo {title} {{Retrospectively modeling the effects of increased
  global vaccine sharing on the COVID-19 pandemic}},\ }\href
  {https://doi.org/10.1038/s41591-022-02064-y} {\bibfield  {journal} {\bibinfo
  {journal} {Nature Medicine}\ }\textbf {\bibinfo {volume} {28}},\ \bibinfo
  {pages} {2416} (\bibinfo {year} {2022})}\BibitemShut {NoStop}%
\bibitem [{\citenamefont {Hill}\ \emph {et~al.}(2022)\citenamefont {Hill},
  \citenamefont {Prosser}, \citenamefont {Ferguson}, \citenamefont {Kaler},
  \citenamefont {Green}, \citenamefont {Keeling},\ and\ \citenamefont
  {Tildesley}}]{Hill2022}%
  \BibitemOpen
  \bibfield  {author} {\bibinfo {author} {\bibfnamefont {E.~M.}\ \bibnamefont
  {Hill}}, \bibinfo {author} {\bibfnamefont {N.~S.}\ \bibnamefont {Prosser}},
  \bibinfo {author} {\bibfnamefont {E.}~\bibnamefont {Ferguson}}, \bibinfo
  {author} {\bibfnamefont {J.}~\bibnamefont {Kaler}}, \bibinfo {author}
  {\bibfnamefont {M.~J.}\ \bibnamefont {Green}}, \bibinfo {author}
  {\bibfnamefont {M.~J.}\ \bibnamefont {Keeling}},\ and\ \bibinfo {author}
  {\bibfnamefont {M.~J.}\ \bibnamefont {Tildesley}},\ }\bibfield  {title}
  {\bibinfo {title} {{Modelling livestock infectious disease control policy
  under differing social perspectives on vaccination behaviour}},\ }\href
  {https://doi.org/10.1371/journal.pcbi.1010235} {\bibfield  {journal}
  {\bibinfo  {journal} {PLOS Computational Biology}\ }\textbf {\bibinfo
  {volume} {18}},\ \bibinfo {pages} {e1010235} (\bibinfo {year}
  {2022})}\BibitemShut {NoStop}%
\bibitem [{\citenamefont {Keeling}\ \emph {et~al.}(2023)\citenamefont
  {Keeling}, \citenamefont {Moore}, \citenamefont {Penman},\ and\ \citenamefont
  {Hill}}]{Keeling2023}%
  \BibitemOpen
  \bibfield  {author} {\bibinfo {author} {\bibfnamefont {M.~J.}\ \bibnamefont
  {Keeling}}, \bibinfo {author} {\bibfnamefont {S.}~\bibnamefont {Moore}},
  \bibinfo {author} {\bibfnamefont {B.~S.}\ \bibnamefont {Penman}},\ and\
  \bibinfo {author} {\bibfnamefont {E.~M.}\ \bibnamefont {Hill}},\ }\bibfield
  {title} {\bibinfo {title} {{The impacts of SARS-CoV-2 vaccine dose separation
  and targeting on the COVID-19 epidemic in England}},\ }\href
  {https://doi.org/10.1038/s41467-023-35943-0} {\bibfield  {journal} {\bibinfo
  {journal} {Nature Communications}\ }\textbf {\bibinfo {volume} {14}},\
  \bibinfo {pages} {1} (\bibinfo {year} {2023})}\BibitemShut {NoStop}%
\bibitem [{\citenamefont {Kucharski}\ \emph {et~al.}(2020)\citenamefont
  {Kucharski}, \citenamefont {Klepac}, \citenamefont {Conlan}, \citenamefont
  {Kissler}, \citenamefont {Tang}, \citenamefont {Fry}, \citenamefont {Gog},
  \citenamefont {Edmunds}, \citenamefont {Emery}, \citenamefont {Medley},
  \citenamefont {Munday}, \citenamefont {Russell}, \citenamefont {Leclerc},
  \citenamefont {Diamond}, \citenamefont {Procter}, \citenamefont {Gimma},
  \citenamefont {Sun}, \citenamefont {Gibbs}, \citenamefont {Rosello},
  \citenamefont {van Zandvoort}, \citenamefont {Hu{\'{e}}}, \citenamefont
  {Meakin}, \citenamefont {Deol}, \citenamefont {Knight}, \citenamefont
  {Jombart}, \citenamefont {Foss}, \citenamefont {Bosse}, \citenamefont
  {Atkins}, \citenamefont {Quilty}, \citenamefont {Lowe}, \citenamefont {Prem},
  \citenamefont {Flasche}, \citenamefont {Pearson}, \citenamefont {Houben},
  \citenamefont {Nightingale}, \citenamefont {Endo}, \citenamefont {Tully},
  \citenamefont {Liu}, \citenamefont {Villabona-Arenas}, \citenamefont
  {O'Reilly}, \citenamefont {Funk}, \citenamefont {Eggo}, \citenamefont {Jit},
  \citenamefont {Rees}, \citenamefont {Hellewell}, \citenamefont {Clifford},
  \citenamefont {Jarvis}, \citenamefont {Abbott}, \citenamefont {Auzenbergs},
  \citenamefont {Davies},\ and\ \citenamefont {Simons}}]{Kucharski2020}%
  \BibitemOpen
  \bibfield  {author} {\bibinfo {author} {\bibfnamefont {A.~J.}\ \bibnamefont
  {Kucharski}}, \bibinfo {author} {\bibfnamefont {P.}~\bibnamefont {Klepac}},
  \bibinfo {author} {\bibfnamefont {A.~J.}\ \bibnamefont {Conlan}}, \bibinfo
  {author} {\bibfnamefont {S.~M.}\ \bibnamefont {Kissler}}, \bibinfo {author}
  {\bibfnamefont {M.~L.}\ \bibnamefont {Tang}}, \bibinfo {author}
  {\bibfnamefont {H.}~\bibnamefont {Fry}}, \bibinfo {author} {\bibfnamefont
  {J.~R.}\ \bibnamefont {Gog}}, \bibinfo {author} {\bibfnamefont {W.~J.}\
  \bibnamefont {Edmunds}}, \bibinfo {author} {\bibfnamefont {J.~C.}\
  \bibnamefont {Emery}}, \bibinfo {author} {\bibfnamefont {G.}~\bibnamefont
  {Medley}}, \bibinfo {author} {\bibfnamefont {J.~D.}\ \bibnamefont {Munday}},
  \bibinfo {author} {\bibfnamefont {T.~W.}\ \bibnamefont {Russell}}, \bibinfo
  {author} {\bibfnamefont {Q.~J.}\ \bibnamefont {Leclerc}}, \bibinfo {author}
  {\bibfnamefont {C.}~\bibnamefont {Diamond}}, \bibinfo {author} {\bibfnamefont
  {S.~R.}\ \bibnamefont {Procter}}, \bibinfo {author} {\bibfnamefont
  {A.}~\bibnamefont {Gimma}}, \bibinfo {author} {\bibfnamefont {F.~Y.}\
  \bibnamefont {Sun}}, \bibinfo {author} {\bibfnamefont {H.~P.}\ \bibnamefont
  {Gibbs}}, \bibinfo {author} {\bibfnamefont {A.}~\bibnamefont {Rosello}},
  \bibinfo {author} {\bibfnamefont {K.}~\bibnamefont {van Zandvoort}}, \bibinfo
  {author} {\bibfnamefont {S.}~\bibnamefont {Hu{\'{e}}}}, \bibinfo {author}
  {\bibfnamefont {S.~R.}\ \bibnamefont {Meakin}}, \bibinfo {author}
  {\bibfnamefont {A.~K.}\ \bibnamefont {Deol}}, \bibinfo {author}
  {\bibfnamefont {G.}~\bibnamefont {Knight}}, \bibinfo {author} {\bibfnamefont
  {T.}~\bibnamefont {Jombart}}, \bibinfo {author} {\bibfnamefont {A.~M.}\
  \bibnamefont {Foss}}, \bibinfo {author} {\bibfnamefont {N.~I.}\ \bibnamefont
  {Bosse}}, \bibinfo {author} {\bibfnamefont {K.~E.}\ \bibnamefont {Atkins}},
  \bibinfo {author} {\bibfnamefont {B.~J.}\ \bibnamefont {Quilty}}, \bibinfo
  {author} {\bibfnamefont {R.}~\bibnamefont {Lowe}}, \bibinfo {author}
  {\bibfnamefont {K.}~\bibnamefont {Prem}}, \bibinfo {author} {\bibfnamefont
  {S.}~\bibnamefont {Flasche}}, \bibinfo {author} {\bibfnamefont {C.~A.}\
  \bibnamefont {Pearson}}, \bibinfo {author} {\bibfnamefont {R.~M.}\
  \bibnamefont {Houben}}, \bibinfo {author} {\bibfnamefont {E.~S.}\
  \bibnamefont {Nightingale}}, \bibinfo {author} {\bibfnamefont
  {A.}~\bibnamefont {Endo}}, \bibinfo {author} {\bibfnamefont {D.~C.}\
  \bibnamefont {Tully}}, \bibinfo {author} {\bibfnamefont {Y.}~\bibnamefont
  {Liu}}, \bibinfo {author} {\bibfnamefont {J.}~\bibnamefont
  {Villabona-Arenas}}, \bibinfo {author} {\bibfnamefont {K.}~\bibnamefont
  {O'Reilly}}, \bibinfo {author} {\bibfnamefont {S.}~\bibnamefont {Funk}},
  \bibinfo {author} {\bibfnamefont {R.~M.}\ \bibnamefont {Eggo}}, \bibinfo
  {author} {\bibfnamefont {M.}~\bibnamefont {Jit}}, \bibinfo {author}
  {\bibfnamefont {E.~M.}\ \bibnamefont {Rees}}, \bibinfo {author}
  {\bibfnamefont {J.}~\bibnamefont {Hellewell}}, \bibinfo {author}
  {\bibfnamefont {S.}~\bibnamefont {Clifford}}, \bibinfo {author}
  {\bibfnamefont {C.~I.}\ \bibnamefont {Jarvis}}, \bibinfo {author}
  {\bibfnamefont {S.}~\bibnamefont {Abbott}}, \bibinfo {author} {\bibfnamefont
  {M.}~\bibnamefont {Auzenbergs}}, \bibinfo {author} {\bibfnamefont {N.~G.}\
  \bibnamefont {Davies}},\ and\ \bibinfo {author} {\bibfnamefont
  {D.}~\bibnamefont {Simons}},\ }\bibfield  {title} {\bibinfo {title}
  {{Effectiveness of isolation, testing, contact tracing, and physical
  distancing on reducing transmission of SARS-CoV-2 in different settings: a
  mathematical modelling study}},\ }\href
  {https://doi.org/10.1016/S1473-3099(20)30457-6} {\bibfield  {journal}
  {\bibinfo  {journal} {The Lancet Infectious Diseases}\ }\textbf {\bibinfo
  {volume} {20}},\ \bibinfo {pages} {1151} (\bibinfo {year}
  {2020})}\BibitemShut {NoStop}%
\bibitem [{\citenamefont {Piguillem}\ and\ \citenamefont
  {Shi}(2020)}]{Piguillem2020}%
  \BibitemOpen
  \bibfield  {author} {\bibinfo {author} {\bibfnamefont {F.}~\bibnamefont
  {Piguillem}}\ and\ \bibinfo {author} {\bibfnamefont {L.}~\bibnamefont
  {Shi}},\ }\href@noop {} {\emph {\bibinfo {title} {{Optimal COVID-19
  Quarantine and Testing Policies}}}},\ \bibinfo {type} {EIEF Working Paper}\
  \bibinfo {number} {20/04}\ (\bibinfo {year} {2020})\BibitemShut {NoStop}%
\bibitem [{\citenamefont {Schnyder}\ \emph
  {et~al.}(2023{\natexlab{b}})\citenamefont {Schnyder}, \citenamefont {Molina},
  \citenamefont {Yamamoto},\ and\ \citenamefont {Turner}}]{Schnyder2023}%
  \BibitemOpen
  \bibfield  {author} {\bibinfo {author} {\bibfnamefont {S.~K.}\ \bibnamefont
  {Schnyder}}, \bibinfo {author} {\bibfnamefont {J.~J.}\ \bibnamefont
  {Molina}}, \bibinfo {author} {\bibfnamefont {R.}~\bibnamefont {Yamamoto}},\
  and\ \bibinfo {author} {\bibfnamefont {M.~S.}\ \bibnamefont {Turner}},\
  }\bibfield  {title} {\bibinfo {title} {{Rational social distancing in
  epidemics with uncertain vaccination timing}},\ }\href
  {https://doi.org/10.1371/journal.pone.0288963} {\bibfield  {journal}
  {\bibinfo  {journal} {PLOS ONE}\ }\textbf {\bibinfo {volume} {18}},\ \bibinfo
  {pages} {e0288963} (\bibinfo {year} {2023}{\natexlab{b}})},\ \Eprint
  {https://arxiv.org/abs/2305.13618} {arXiv:2305.13618} \BibitemShut {NoStop}%
\bibitem [{\citenamefont {Bensoussan}\ \emph {et~al.}(2013)\citenamefont
  {Bensoussan}, \citenamefont {Frehse},\ and\ \citenamefont
  {Yam}}]{Bensoussan}%
  \BibitemOpen
  \bibfield  {author} {\bibinfo {author} {\bibfnamefont {A.}~\bibnamefont
  {Bensoussan}}, \bibinfo {author} {\bibfnamefont {J.}~\bibnamefont {Frehse}},\
  and\ \bibinfo {author} {\bibfnamefont {P.}~\bibnamefont {Yam}},\ }\href
  {https://doi.org/10.1007/978-1-4614-8508-7} {\emph {\bibinfo {title} {Mean
  Field Games and Mean Field Type Control Theory}}}\ (\bibinfo  {publisher}
  {Springer},\ \bibinfo {year} {2013})\BibitemShut {NoStop}%
\bibitem [{\citenamefont {Carmona}\ and\ \citenamefont
  {Delarue}(2018)}]{Carmona}%
  \BibitemOpen
  \bibfield  {author} {\bibinfo {author} {\bibfnamefont {R.}~\bibnamefont
  {Carmona}}\ and\ \bibinfo {author} {\bibfnamefont {F.}~\bibnamefont
  {Delarue}},\ }\href {https://doi.org/10.1007/978-3-319-58920-6} {\emph
  {\bibinfo {title} {Probabilistic Theory of Mean Field Games with Applications
  I}}}\ (\bibinfo  {publisher} {Springer},\ \bibinfo {year} {2018})\BibitemShut
  {NoStop}%
\bibitem [{\citenamefont {Lenhart}\ and\ \citenamefont
  {Workman}(2007)}]{OptimalControlBook}%
  \BibitemOpen
  \bibfield  {author} {\bibinfo {author} {\bibfnamefont {S.}~\bibnamefont
  {Lenhart}}\ and\ \bibinfo {author} {\bibfnamefont {J.}~\bibnamefont
  {Workman}},\ }\href@noop {} {\emph {\bibinfo {title} {Optimal Control Applied
  to Biological Models}}}\ (\bibinfo  {publisher} {Chapman and Hall/CRC},\
  \bibinfo {year} {2007})\BibitemShut {NoStop}%
\bibitem [{\citenamefont {Pontryagin}\ \emph {et~al.}(1986)\citenamefont
  {Pontryagin}, \citenamefont {Boltyanskii}, \citenamefont {Gamkrelidze},\ and\
  \citenamefont {Mishchenko}}]{Pontryagin}%
  \BibitemOpen
  \bibfield  {author} {\bibinfo {author} {\bibfnamefont {L.~S.}\ \bibnamefont
  {Pontryagin}}, \bibinfo {author} {\bibfnamefont {V.}~\bibnamefont
  {Boltyanskii}}, \bibinfo {author} {\bibfnamefont {R.~V.}\ \bibnamefont
  {Gamkrelidze}},\ and\ \bibinfo {author} {\bibfnamefont {E.~F.}\ \bibnamefont
  {Mishchenko}},\ }\href@noop {} {\emph {\bibinfo {title} {{The Mathematical
  Theory of Optimal Processes}}}}\ (\bibinfo  {publisher} {Gordon and Breach
  Science Publishers},\ \bibinfo {year} {1986})\BibitemShut {NoStop}%
\end{thebibliography}%

\end{document}